\documentclass[useAMS,usenatbib]{mn2e}

\usepackage{epsfig}
\usepackage{myaasmacros}
\usepackage{amsmath}
\usepackage{Arial}
\usepackage{color}

\usepackage[normalem]{ulem}

\title[Magnetic fields in colliding galaxies]
{Galactic m\'{e}nage \`{a} trois: Simulating magnetic fields in colliding galaxies}
\author[H. Kotarba, H. Lesch, K. Dolag, T. Naab, P. H. Johansson, J. Donnert \& F. A. Stasyszyn]{H. Kotarba$^{1,2}$\thanks{E-mail:
kotarba@usm.lmu.de}, H. Lesch$^{1}$, K. Dolag$^{1,3}$, T. Naab$^{3,1}$, P. H. Johansson$^{4,5,1}$, \newauthor J. Donnert$^{3}$ \& F. A. Stasyszyn$^{1,3}$\\
$^{1}$University Observatory Munich, Scheinerstr.1, D-81679 Munich, Germany\\
$^{2}$Max Planck Institute for Extraterrestrial Physics, Giessenbachstrasse, D-85748 Garching, Germany\\
$^{3}$Max Planck Institute for Astrophysics, Karl-Schwarzschild-Str. 1, D-85741 Garching, Germany\\
$^{4}$Finnish Centre for  Astronomy with ESO, University of Turku, V\"ais\"al\"antie 20, FI-21500 Piikki\"o, Finland\\
$^{5}$Department of Physics, University of Helsinki, Gustaf H\"allstr\"omin katu 2a, FI-00014 Helsinki, Finland}

\begin{document}

\date{Accepted  \hspace{3mm} Received 25 November 2010 \hspace{3mm} in original form 25 November 2010}

\pagerange{\pageref{firstpage}--\pageref{lastpage}} \pubyear{2010}

\maketitle

\label{firstpage}

\begin{abstract}

We present high resolution simulations of a multiple merger of three disk galaxies including the evolution of magnetic fields performed with the \textsl{N}-body/SPH code $\textsc{Gadget}$. For the first time, we embed the galaxies in a magnetized, low-density medium, thus modeling an ambient IGM. The simulations include radiative cooling and a model for star formation and supernova feedback. Magnetohydrodynamics is followed using the SPH method. The progenitor disks have initial magnetic seed fields in the range of $10^{-9}$ to $10^{-6}$ G and the IGM has initial fields of $10^{-12}$ to $10^{-9}$ G. The simulations are compared to a run excluding magnetic fields. We show that the propagation of interaction-driven shocks depends significantly on the initial magnetic field strength. The shocks propagate faster in simulations with stronger initial field, suggesting that the shocks are supported by magnetic pressure. The Mach numbers of the shocks range from approximately $M=1.5$ for the non-magnetized case up to $M=6$ for the highest initial magnetization, resulting in higher temperatures of the shock heated IGM gas. The magnetic field in the system saturates rapidly after the mergers at $\sim 10^{-6}$ G within the galaxies and $\sim 10^{-8}$ G in the IGM independent of the initial value. These field strengths agree with observed values and correspond to the equipartition value of the magnetic pressure with the turbulent pressure in the system. We also present synthetic radio and polarization maps for different phases of the evolution showing that shocks driven by the interaction produce a high amount of polarized emission. These idealized simulations indicate that magnetic fields play an important role for the hydrodynamics of the IGM during galactic interactions. We also show that even weak seed fields are efficiently strengthened during multiple galactic mergers. This interaction driven amplification might have been a key process for the magnetization of the Universe.
\end{abstract}


\section{Introduction}

Radio observations have revealed that most late type galaxies in the local Universe - isolated grand design spirals, irregulars and dwarf galaxies - are permeated by magnetic fields (\citealp{BeckKrauseKlein1985,HummelBeck1995,BeckHoernes1996,Chyzy2007,Vollmer2010}). The field strengths in all of these objects do not vary by more than one order of magnitude from a few $\mu$G in dwarfs (e.g. \citealp{Chyzy2003}) to 30 $\mu$G in the star-forming regions of grand-design spiral galaxies (\citealp{Fletcher2004}). Magnetic fields have also been observed at redshifts up to $z\approx2$ in damped Ly-$\alpha$ systems. These systems, which might be interpreted as large progenitors of present-day galaxies (e.g. \citealp{Wolfe1995,Wolfe2005review}), seem to host magnetic fields of similar strength as local late-type galaxies (e.g. \citealp{Bernet2008} and references therein).

These observations invite the question about the origin and the evolution of the magnetic fields in the early universe. Different scenarios have been suggested: \citet{Lesch&Chiba1995} have shown analytically that strong magnetic fields in high redshift objects can be explained by the combined action of an evolving protogalactic fluctuation and electrodynamic processes providing magnetic seed fields (i.e. battery processes). \citet{Wang&Abel2009} performed numerical simulations of the formation of disc galaxies within an collapsing halo imposing a uniform initial magnetic field of $10^{-9}$ G. The initial field grew by three orders of magnitude within approximately 500 Myr of the evolution. The amplification might be due to the combined effects of magnetic field compression during the collapse and amplification of the uniform initial field by differential rotation as studied also in \citet{me2009}. These studies indicate, that the growth of magnetic fields might be a natural part of the galaxy formation process.

A key ingredient in galaxy formation studies, however, is the consideration of galaxy interactions. Within the standard cold dark matter (CDM) models present-day galaxies have undergone several major and minor mergers at earlier epochs of the universe, and thereafter continued accreting gas and smaller galactic subunits (\citealp{White&Rees1978,White&Frenk1991}).
Interactions of galaxies change their dynamics drastically (\citealp{Toomre&Toomre1972, Barnes1992,Hernquist1994, Barnes1999,Naab&Burkert2003,Gonzalez-Gracia2006}) as the gravitational potential is varying rapidly during the interaction.  Since the gas component is dissipative and most sensitive to variations of the gravitational potential, it is strongly affected by the interaction and driven towards the galaxy centers, eventually causing bursts of star formation (\citealp{Barnes&Hernquist1992,Mihos&Hernquist1994, Barnes&Hernquist1996, Bekki&Shioya1998,Springel2000,Barnes2002, Bournaud2005,Cox2006,Naab&Jesseit2006,Robertson2006,Cox2008,Hopkins2008,Teyssier2010}). We can crudely estimate the amount of free energy during an interaction of two galactic subunits to be proportional to their relative velocity squared, i.e. $E_\mathrm{free} \sim v_\mathrm{rel}^2$. Obviously some of this energy released during the interaction is converted into thermal energy of hot gas. High energy particles also carry away some of the energy. However, it is reasonable to assume that at least some of this energy is converted into magnetic field energy during the compression of gas and the formation of tidal structures. As the amount of $E_\mathrm{free}$ can be very large during a major merger, the amount of energy converted into magnetic energy might be significant. Moreover, gas which is heated by the interaction and driven into the IGM should carry magnetic energy out of the galactic units, thus magnetizing the IGM. This process should be similar to what was found by \citet{DuboisTeyssier2009} in their study of the formation of dwarf galaxies including magnetic fields and galactic winds.

So far, simulations of interactions and mergers of disk galaxies have been mainly investigated with respect to changes in stellar dynamics, gas flows, star formation (SF) and the formation of central supermassive black holes (\citealp{DiMatteo2005Natur, Springel2005BlackHoles,Springel2005,Robertson2006,Johansson2009BlackHoles,Johansson2009}). However, the dramatic impact of mergers on the gas flows will directly affect the dynamics of the magnetic field of the systems. Since gas and magnetic field are tightly coupled, the magnetic field traces the gas motion and will be strengthened by shocks and gas inflow. A perfect example for the strong coupling of gas and magnetic fields is the interacting system NGC4038/4039 (the ''Antennae galaxies``) which has been recently simulated (\citealp{Karl2010}) by \citet{meAnt2009} including magnetic fields. It was shown that even an initial magnetic field as small as $10^{-9}$ G grows significantly during the interaction of two equal-mass spiral galaxies. The magnetic field strength thereby saturates at a value of $\approx 10$ $\mu$G, in good agreement with observations (\citealp{ChyzyBeck2004}). This saturation value was reached independently of the initial magnetic field strength in the range of $B_0 = 10^{-9}$ - $10^{-4}$ G. We emphasize that \citet{meAnt2009} found the saturation level to correspond to near equipartition between magnetic and turbulent gas pressure, which is in good agreement with theoretical considerations of the turbulent dynamo (e.g. \citealp{Arshakian2009} and references therein, see also section \ref{Pressures}). Furthermore, \citet{meAnt2009} provided synthetic radio maps calculated at time of best match between the simulated gas and stellar distributions and observations. These synthetic radio maps are in convincing morphological agreement with synchrotron observations of the Antennae system and the underlying numerics of the applied \textsl{N}-body/SPH code $\textsc{Gadget}$ showed to be capable of following the evolution of magnetic fields in a highly nonlinear environment.

High resolution simulations of the formation of individual galaxies in a full cosmological context (see e.g. \citealp{Naab2007,Sawala2010,Piontek2010}) including magnetic fields could help us in understanding the processes leading to the magnetization of the Universe. Although it would be worthwile to consider cosmological studies in the long run, the simulations of three interacting galaxies presented in this paper are an additional further step towards a more complete scenario and a natural extension of the previous study presented in \citet{meAnt2009}. With this study we show that the magnetic field growth accompanying a galactic interaction and its saturation at the equipartition level between magnetic and turbulent pressure holds also for a more general setup of interacting galaxies including IGM gas.

The paper is organized as follows: We briefly describe our numerical method in section \ref{NUMERICS}. In section \ref{SETUP}, we present a detailed description of the setup of the three colliding galaxies. The temporal evolution of the simulated systems and particulary the magnetic fields is described in section \ref{SIMULATIONS}. In section \ref{RADIO}, we present synthetic radio and $RM$ maps of our simulated system. Finally, we summarize our results and conclude in section \ref{CONCLUSION}.

\section{Numerical Method}\label{NUMERICS}

The simulations presented here were performed with the $\mathrm{N}$-body/SPH-code $\textsc{Gadget}$ (\citealp{SpringelGadget}). Gravitational interactions between the particles are evaluated with a hierarchical tree method (\citealp{Barnes&Hut1986}). The dynamics of the Lagrangian fluid elements are followed using a SPH formulation which conserves both energy and entropy (\citealp{Springel&Hernquist2002}) including the evolution of magnetic fields which was implemented and tested by \citet{GadgetMHD}. The code has already been used to investigate the evolution of magnetic fields in isolated spiral galaxies (\citealp{me2009}) and during the collision of two equal mass spiral galaxies resembling the nearby Antennae galaxies (\citealp{meAnt2009}).

The simulations presented in this paper have been performed using the standard (direct) magnetic field implementation. We apply the Lorentz force and artificial magnetic dissipation with an artificial magnetic dissipation constant of $\alpha_B=0.5$. Artificial magnetic dissipation is included in order to reduce numerical errors arising from the SPH approximation rather than to capture any physical dissipation correctly (\citealp{GadgetMHD}). Thereby, only a small fraction of the magnetic field within the volume defined by the SPH particle is allowed to leave this volume within the local dynamical time (defined via the local signal velocity). This limitation ensures that even strong shocks are well captured. Yet, of course, the artificial dissipation also leads to an effective slow diffusion of the magnetic field. This can be seen in the simulations of isolated disk galaxies presented in \citet{meAnt2009} (Fig. 3 in their paper, the yellow and red lines). However, the value of $\alpha_B$ is chosen such that on the one hand the numerical errors are efficiently reduced, and on the other hand, the magnetic diffusion is preferably low. Hence, the numerical dissipation does not reflect the true physical dissipation. Physical magnetic dissipation arises either due to electric conductivity, in which case it is very small and thus negligible, or due to turbulent diffusion, which can be significantly higher. However, turbulent diffusion would have to be modeled within the simulations, because it reflects processes on sub-resolution scales. We do not include any sub-resolution turbulent diffusion model in our simulations.

In addition to \citet{meAnt2009}, we now also apply the subtraction of the effect of numerical magnetic divergence (the ``divergence force'') in the momentum equation as suggested by \citet{Borve2001} (see \citealp{GadgetMHD} for more details). However, as a refinement of the method presented in \citet{GadgetMHD}, we define a threshold for the divergence force subtraction: Whenever the correction becomes larger than the half of the current Lorentz force, it is limited to that level. In this way we avoid situations in which the divergence force could become the main source of acceleration and thus instabilities due to temporal high numerical divergence, e.g. during strong interactions (see also section \ref{globalEvolutionDIVB}). This limitation of the divergence force also helps to maintain energy conservation (for details see Stasyszyn \& Dolag (2011), in preparation). The basic method of divergence force subtraction is not conservative, however, by limiting the correction we reduce possible transfer of energy associated with the numerical divergence to kinetic energy.

Similarly to \citet{meAnt2009}, we apply radiative cooling, star formation and the associated supernova feedback, but exclude explicit supernova-driven galactic winds, following the sub-resolution multiphase model developed by \citet{Springel&Hernquist2003}, in which the ISM is treated as a two-phase medium (\citealp{McKee1977}, \citealp{Johansson&Efstathiou2006}).

The implementation used in this paper has been tested in detail (\citealp{SpringelGadget1,SpringelGadget,Springel2005,GadgetMHD}). We refer the reader to these studies for further details on the applied method. Detailed discussions on the numerical divergence of the magnetic field ($\nabla\cdot\textbf{B}$) in SPH simulations of isolated and interacting galaxies can be found in \citet{me2009} and \citet{meAnt2009}.

\section{Setup}\label{SETUP}

\subsection{Galaxies}

The initial conditions for the spiral galaxies are produced using the method described by \citet{Springel2005} which is based on \citet{Hernquist1993}. The galaxies consist of a cold dark matter halo, a rotationally supported exponential stellar disk, an exponential gas disk and a stellar bulge component. The profiles of the halo and bulge are based on \citet{Hernquist1990}. The halo, stellar disk and bulge particles are collisionless $\mathrm{N}$-body particles. The gas is represented by SPH particles.

The parameters used for the initial setup of the three identical galaxies can be found in Table \ref{tab1}. The magnetic field in each disc is set to $B_x=B_{0\mathrm{,disk}}$ and $B_y=B_z=0$ with the $z$-axis being the axis of rotation. The particle numbers and fixed gravitational softening lengths $\epsilon$ for each galaxy can be found in Table \ref{tab2}. The minimum SPH smoothing length for the gas particles is thereby $h_\mathrm{SPH}^\mathrm{min}=1.0\epsilon=25$ pc$/h$, with $h=0.71$ being the Hubble constant. This choice of parameters results in particle masses of $m_\mathrm{gas}=m_\mathrm{disk}=m_\mathrm{bulge}\approx1.19\cdot10^5 M_\odot$/$h$ and $m_\mathrm{halo}\approx21.43\cdot10^5 M_\odot$/$h$, respectively, with the Hubble constant $h=0.71$. More details on the properties, evolution and stability of the disks evolved in isolation can be found in \citet{me2009,meAnt2009}.

\begin{table}
{\footnotesize \begin{tabular}{p{4cm} p{1cm} p{2cm}}
\hline
\multicolumn{3}{c}{\textsc{Disk parameters}}\\\hline\hline
  total mass & $M_\mathrm{tot}$   &  $1.34\times 10^{12}M_\odot$    \\
  disk mass  & $M_\mathrm{disk}$  &  0.075 $M_\mathrm{tot}$   \\
  bulge mass & $M_\mathrm{bulge}$ &  0.025 $M_\mathrm{tot}$ \\
  mass of the gas disk & $M_\mathrm{gas}$ & 0.2 $M_\mathrm{disk}$  \\
  exponential disk scale length & $l_D$   & 8.44 kpc\\
  scale height of the disk      & $h_D$     & 0.2 $l_D$ \\
  bulge scale length            & $l_B$   & 0.2 $l_D$ \\
  spin parameter              & $\lambda$ & 0.1 \\
  virial velocity of the halo & $v_\mathrm{vir}$  & 160 km s$^{-1}$ \\
  half mass radius            & $R_\mathrm{half}$ & $\approx$12 kpc\\
  half mass circular velocity & $v_\mathrm{half}$ & $\approx$249 km s$^{-1}$ \\
  half mass rotation period   & $T_\mathrm{half}$ & $\approx$295 Myr \\
  initial temperature         & $T_\mathrm{disk}$  & $\approx$10 000 K \\
  initial magnetic field      & $B_0$ & $0$ - $10^{-6}$ G \\\hline
  \multicolumn{3}{c}{\textsc{Multi-Phase model parameters}}\\\hline\hline
  gas consumption timescale   & $t_\mathrm{MP}$ & 8.4 Gyr \\
  mass fraction of massive stars & $\beta_\mathrm{MP}$ & 0.1 \\
  evaporation parameter & $A_0$ & 4000 \\
  effective SN temperature & $T_\mathrm{SN}$ & $4\times10^8$ K \\
  cold cloud temperature & $T_\mathrm{CC}$ & 1000 K \\\hline
\end{tabular}}
  \caption{Parameters of initial disk setup}
  \label{tab1}
\end{table}

\begin{table}
{\footnotesize \begin{tabular}{p{2.5cm} p{2cm} p{2.5cm}}
\hline
Component & initial particle number & fixed gravitational softening length $\epsilon$ [pc]$^{\rm{a}}$ \\\hline\hline
  Dark Matter         &  $4.0\times10^5$  & 110/$h$  \\
  Disk - stars        &  $4.8\times10^5$  & 25/$h$  \\
  Bulge - stars       &  $2.0\times10^5$  & 25/$h$  \\
  Gas                 &  $1.2\times10^5$  & 25/$h$  \\
  Total               &  $1.2\times10^6$  & -       \\\hline
  \multicolumn{3}{p{8cm}}{\scriptsize{(a) The Hubble constant is assumed to be $h=0.71$ in this paper.}} \\
\end{tabular}}
  \caption{Particle numbers and softening lengths}
  \label{tab2}
\end{table}

\subsection{Orbits}

We want to study a general case of a galactic interaction. Thus, the initial orbital setup of the three colliding galaxies has been chosen arbitrarily without the aim of matching a particular observed system. In creating the initial conditions, the galaxies (including their magnetic field) are first rotated with respect to the plane of sky ($xy$-plane) around the $x$-, $y$- and $z$-axes by the angles $\theta$, $\psi$ and $\phi$, respectively. In order to guarantee a final merger of the three galaxies, two of them (galaxy 1 (G1) and galaxy 2 (G2)) are set on a nearly parabolic Keplerian two-body orbit with an initial separation of the centers of mass of $r_\mathrm{sep}^a$ and a pericenter distance of $r_\mathrm{p}^a$. Then, the third galaxy (G3) and the center of mass of the combined system G1 and G2 are set on a nearly parabolic orbit with $r_\mathrm{sep}^b$ and $r_\mathrm{p}^b$. The values of the rotation angles, the initial separations, the pericenter distances and the initial velocities can be found in Table \ref{tab3}.

\begin{table}
{\footnotesize \begin{tabular}{p{3.6cm} p{1cm} p {1cm} p{1cm}}
\hline
  initial separation [kpc/$h$]$^{\rm{a}}$  & \multicolumn{2}{r}{$r_\mathrm{sep}^a =$} & \multicolumn{1}{r}{160}\\
                                           & \multicolumn{2}{r}{$r_\mathrm{sep}^b =$} & \multicolumn{1}{r}{320}\\
  pericenter distance [kpc/$h$]$^{\rm{a}}$ & \multicolumn{2}{r}{$r_\mathrm{p}^a =$}   & \multicolumn{1}{r}{12} \\
                                           & \multicolumn{2}{r}{$r_\mathrm{p}^b =$}   & \multicolumn{1}{r}{24} \\\hline\hline
                                & \hspace{5mm} G1          & \hspace{5mm}  G2        & \hspace{5mm} G3 \\
  rotation around $x$ ($\theta$)& \multicolumn{1}{r}{45}   & \multicolumn{1}{r}{90}  & \multicolumn{1}{r}{30} \\
  rotation around $y$ ($\psi$)  & \multicolumn{1}{r}{45}   & \multicolumn{1}{r}{90}  & \multicolumn{1}{r}{0}  \\
  rotation around $z$  ($\phi$) & \multicolumn{1}{r}{0}    & \multicolumn{1}{r}{0}   & \multicolumn{1}{r}{0}  \\
  initial $v_x$ [km s$^{-1}$]   & \multicolumn{1}{r}{-145} & \multicolumn{1}{r}{73}  & \multicolumn{1}{r}{18} \\
  initial $v_y$ [km s$^{-1}$]   & \multicolumn{1}{r}{-157} & \multicolumn{1}{r}{-95} & \multicolumn{1}{r}{63} \\\hline
  \multicolumn{4}{p{8cm}}{\scriptsize{(a) The Hubble constant is assumed to be $h=0.71$ in this paper.}} \\
\end{tabular}}
  \caption{Collision setup parameters}\label{tab3}
\end{table}

\subsection{IGM}

As an extension compared to our previous study in \citet{meAnt2009}, we now also include an ambient IGM surrounding the galaxies in order to be able to realistically study the magnetic field evolution in this IGM. The IGM is set up by placing additional gas particles in a regular hexagonal close-packed lattice (hcp) arrangement. Close-packed lattice arrangements have been shown to be more stable for the particles than simple grid arrangements (see \citealp{PriceBate2007} and references therein). The particle mass is again $m_\mathrm{IGM}=m_\mathrm{gas}\approx1.19\cdot10^5 M_\odot$/$h$. The volume filled by the IGM is $500\times700\times300$ (kpc/$h$)$^3$ and centered at the center-of-mass of the initial setup of the discs. The density is $\rho_\mathrm{IGM}=10^{-29}$ g cm$^{-3}$ (equivalent to $\approx 6\cdot10^{-6}$ H-atoms cm$^{-3}$), resulting in a particle number of $N_\mathrm{IGM}=43\times70\times31=93310$.

The initial morphology of our galaxies corresponds to fully evolved disk galaxies residing in their dark matter haloes, hence, we assume that the IGM in each scenario is already virialized. However, for simplicity, we assume a common temperature for the IGM, which we set to the virial temperature at the virial radius of the simulated galactic haloes:

\begin{equation}
T_\mathrm{vir}=\frac{\langle v^2\rangle\mu m_p}{3k_B}=\frac{GM\mu m_p}{3 r_\mathrm{vir}k_B}=T_\mathrm{IGM}\approx 6\cdot10^5 \mbox{ K,}
\end{equation}

where $\mu\approx0.588$ is the molecular weight (for fully ionized gas of primordial composition), $m_p$ the proton mass, $k_B$ the Boltzmann constant, $G$ the gravitational constant, $M$ the mass of the halo and $r_\mathrm{vir}$ its virial radius. We assume the initial magnetic field of the IGM gas to be directed in the $x$ direction (where the $x-y$-plane is the orbital plane) with $B_x=B_{0\mathrm{,IGM}}$. Thus, altogether, the galaxies are permeated by a homogeneous magnetic field lying in the planes of the disks, and the IGM hosts a homogeneous field lying in the orbital plane.

The addition of an ambient IGM has a further, numerical advantage: As the field is not dropping to zero at the disk edges, spurious calculations of the numerical divergence $h_\mathrm{SPH}\nabla\cdot\textbf{B}/|\textbf{B}|$ (where $h_\mathrm{SPH}$ is the SPH smoothing length) are avoided, resulting in higher numerical stability.

\subsection{Magnetic fields}

A detailed study of the cosmological seeding and evolution of magnetic fields in the early universe is beyond the scope of this paper. Rather, we want to gain better understanding of how magnetic fields might have evolved during the epoch of galaxy formation by considering three different scenarios of the magnetization of the colliding galaxies in the local universe, and, additionally, a further scenario excluding magnetic fields for comparison.

The initial magnetic field strengths for each scenario are summarized in Table \ref{tab0}. With the first scenario (G6-IGM9), we aim to simulate a present-day galactic merger. Hence, within this scenario, we assume that the galaxies already host an initial magnetic field of $B_{0\mathrm{,disk}}=10^{-6}$ G, and that the IGM is interspersed with an initial magnetic field of $B_{0\mathrm{,IGM}}=10^{-9}$ G. With our second scenario (G9-IGM9), we want to study the general situation of a common magnetic field strength in the IGM and the galaxies. Within this scenario, the galaxies as well as the IGM host an initial magnetic field of $10^{-9}$ G. In the third scenario (G9-IGM12), we assume that the magnetic field strengths are weaker by roughly three orders of magnitude than today, i.e. $B_{0\mathrm{,disk}}=10^{-9}$ G, and  $B_{0\mathrm{,IGM}}=10^{-12}$ G. In the last scenario (G0-IGM0), we exclude any magnetic field. For simplicity, all initial magnetic fields are assumed to be homogeneous with only one non-vanishing component at the beginning of the simulations (which lies in the plane of the disks for the galaxies, and in the orbital plane for the IGM). The choice of the initial configuration is not important, as the timescales of the simulations are much longer than e.g. the turbulent timescales of the gas. Moreover, it takes more than 0.5 Gyr until the first encounter between the galaxies and the magnetic field within the galaxies has enough time to redistribute and form a realistic configuration prior to the first encounter (see also \citealp{me2009, meAnt2009}).

\begin{table}
{\footnotesize \begin{tabular}{p{2cm} p{2cm} p{2cm}}
\hline
 Scenario & $B_{0\mathrm{,disk}}$ [G] & $B_{0\mathrm{,IGM}}$ [G] \\\hline\hline
  G6-IGM9     &  $10^{-6}$   & $10^{-9}$  \\
  G9-IGM9     &  $10^{-9}$   & $10^{-9}$  \\
  G9-IGM12    &  $10^{-9}$   & $10^{-12}$ \\
  G0-IGM0     &     -        &     -      \\\hline
\end{tabular}}
  \caption{Initial uniform magnetic field strengths for the different scenarios in the plane of the disks ($B_{0\mathrm{,disk}}$) and in the $x$-direction of the orbital plane ($B_{0\mathrm{,IGM}}$), respectively.}
  \label{tab0}
\end{table}

Finally, we let the system evolve for 200 Myr in order to allow possible numerical discontinuities introduced by the setup to relax before considering its physical properties.

\section{Simulations}\label{SIMULATIONS}

\subsection{Morphological evolution}\label{morphological_evol}

\begin{figure*}
\begin{center}
  \epsfig{file=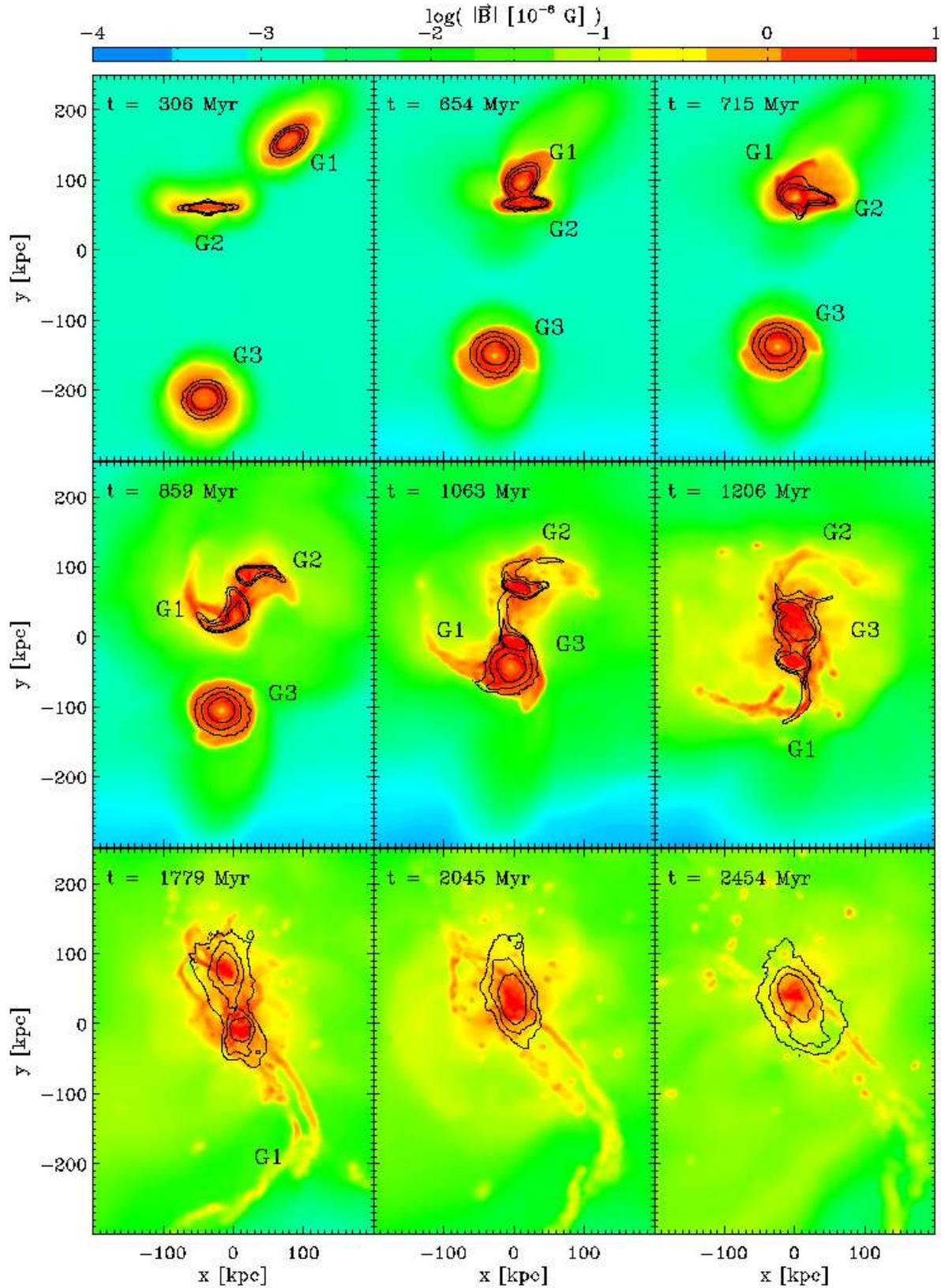, width=0.9\textwidth}
\end{center}
  \caption{The evolution of the magnetic field as a function of time for the G6-IGM9 scenario. Colours visualize the mean line-of-sight total magnetic field $|\mathbf{B}|$ (in units of $10^{-6}$ G) and contours correspond to stellar surface density $\Sigma_\mathrm{stars}$. The contour levels are 10, 20 and 50 $M_\odot$ pc$^{-2}$.
  \label{evolution_bfld_G6-IGM9}}
\end{figure*}

\begin{figure*}
\begin{center}
  \epsfig{file=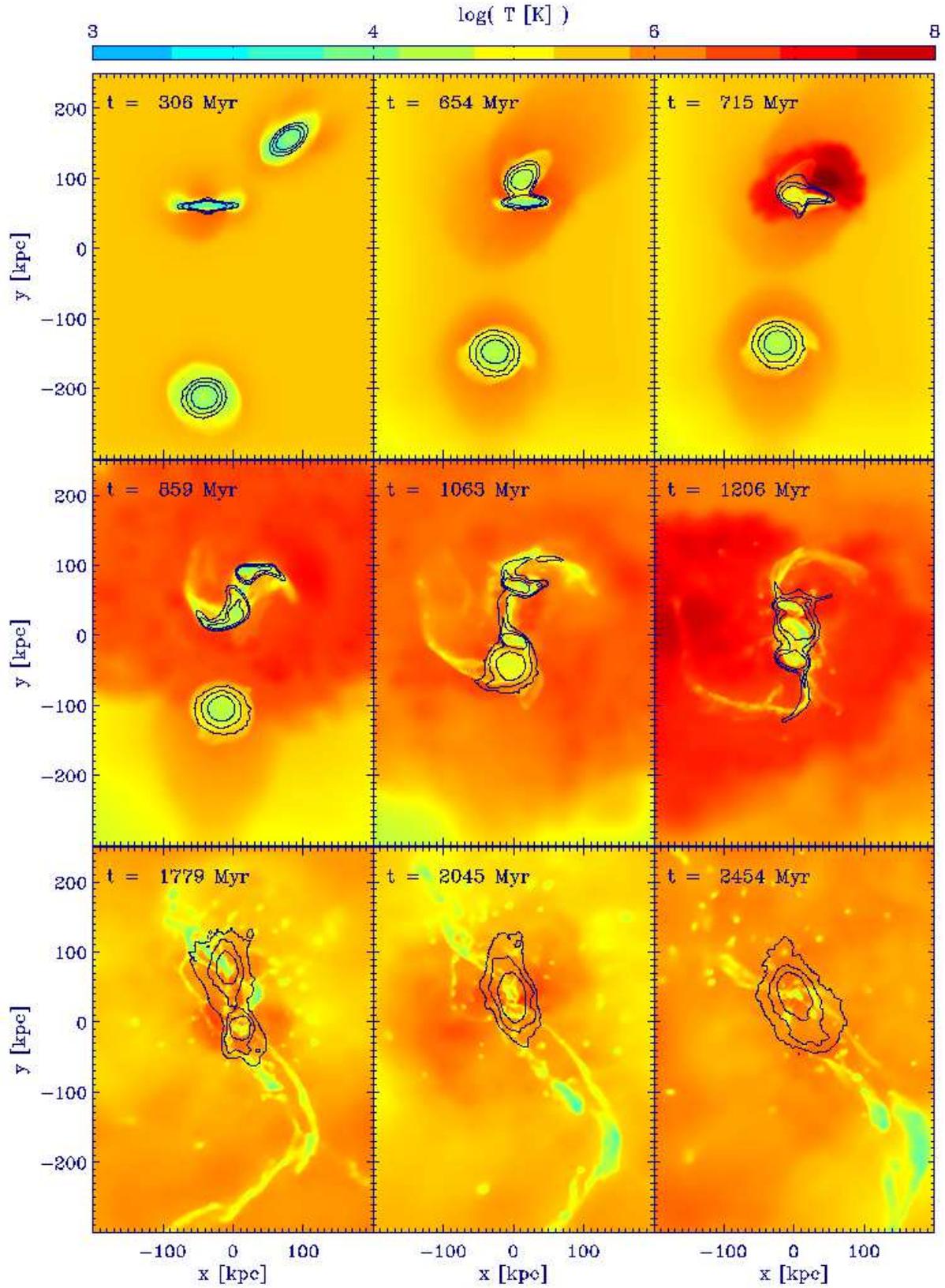, width=0.9\textwidth}
\end{center}
  \caption{The evolution of the temperature as a function of time for the G6-IGM9 scenario. Colours visualize the mean line-of-sight temperature $T$ (in K). Contours correspond to stellar surface density as in Fig. \ref{evolution_bfld_G6-IGM9}.
  \label{evolution_temp_G6-IGM9}}
\end{figure*}

\begin{figure*}
\begin{center}
  \epsfig{file=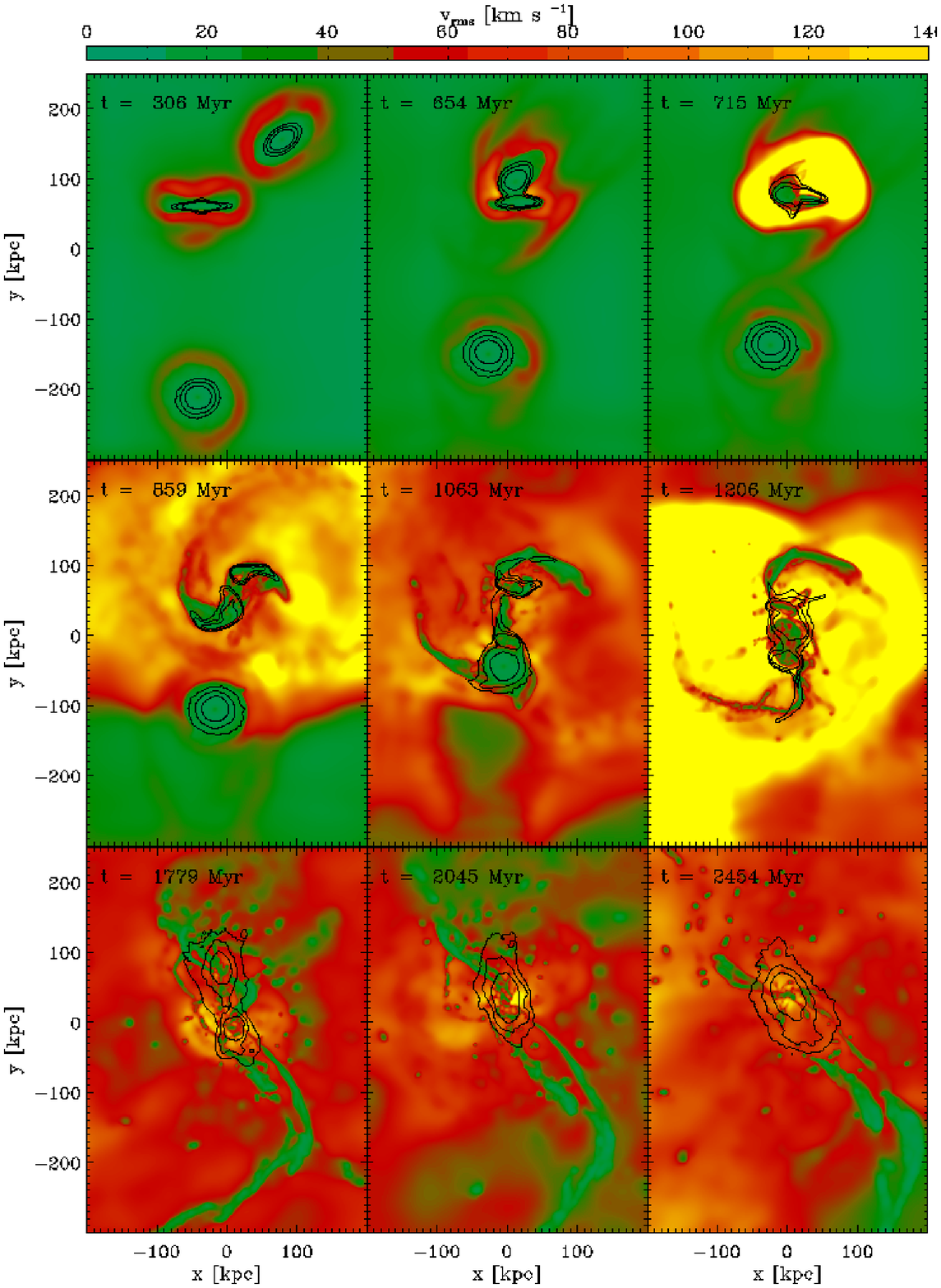, width=0.9\textwidth}
\end{center}
  \caption{The evolution of the rms velocity as a function of time for the G6-IGM9 scenario. Colours visualize the mean line-of-sight rms velocity $v_\mathrm{rms}$ (in km s$^{-1}$). Contours correspond to stellar surface density as in Fig. \ref{evolution_bfld_G6-IGM9}.
  \label{evolution_vrms_G6-IGM9}}
\end{figure*}

Figs. \ref{evolution_bfld_G6-IGM9}, \ref{evolution_temp_G6-IGM9} and \ref{evolution_vrms_G6-IGM9} show the mean line-of-sight total magnetic field $|\mathbf{B}|$, the mean line-of-sight temperature $T$ and the mean line-of-sight rms velocity $v_\mathrm{rms}$, respectively, at nine different time steps for the G6-IGM9 scenario (i.e. the present-day merger)\footnote{See http://www.usm.uni-muenchen.de/people/kotarba/public.html for the corresponding movies.}. To generate the images, the particle data in the 500$^3$ (kpc/$h$)$^3$ volume was binned to a 512$^2$ image using the code \textsc{P-Smac2} (Donnert et al., in preparation), which applies the gather approximation (see \citealp{Dolag2005}). In each case, the color maps are overlaid with contours of the stellar surface density $\Sigma_\ast$ in order to indicate the stellar morphology of the system. The stellar density was binned using the TSC procedure (Triangular Shaped Cloud, see e.g. \citealp{Hockney&Eastwood1988}). Here and hereafter, $v_\mathrm{rms}$ of each particle is defined as the rms velocity around the \textit{mean velocity} inside the SPH kernel of the particle (i.e. inside its smoothing length $h_\mathrm{SPH}$). This value is different from the rms velocity around the \textit{central velocity }(i.e. the velocity of the particle considered) inside the kernel. The latter was defined as the ``turbulent'' velocity in \citet{meAnt2009}. However, $v_\mathrm{rms}$ is a more conservative estimator for the local turbulent velocity on the kernel scale. By averaging over all particles inside the kernel to obtain the mean we reduce the influence of SPH sampling noise, which is present on subkernel scales, on the estimate. Hence, we use $v_\mathrm{rms}$ as the estimate of the turbulent velocity in this paper.

At $t=300$ Myr, the initial magnetic field has been redistributed and forms a realistic configuration in each galaxy (upper left panel in Fig. \ref{evolution_bfld_G6-IGM9}). Due to the winding of the initially uniform magnetic field by differential rotation, the magnetic field in each galaxy forms a non-axisymmetric pattern, which can be recognized from the two magnetic arms extending from the galactic disc (best visible at the later times $t=654$ Myr and $t=715$ Myr in G3, see also \citealp{me2009,meAnt2009} for a more detailed discussion of the winding process). As a consequence of their mutual gravitational attraction, G1 and G2 are moving towards each other and collide at $t\approx 0.7$ Gyr. Due to the lower ram pressure within the IGM than within the galaxies, the interaction-driven shocks (which are analyzed in more detail in section \ref{Differences}) are propagating favorably into the IGM, thus heating the IGM gas (upper central and right panels in Fig. \ref{evolution_temp_G6-IGM9}). Simultaneously, the rms velocity is enhanced and the magnetic field is strengthened in the shocked regions (upper right panels in Figs. \ref{evolution_bfld_G6-IGM9} and \ref{evolution_vrms_G6-IGM9}). During the collision, prominent tidal arms are developing (middle left panels in Figs. \ref{evolution_bfld_G6-IGM9} - \ref{evolution_vrms_G6-IGM9}). At $t\approx 1.1$ Gyr a second collision between G1 and G3 takes place, again accompanied by shocks and interaction-driven outflows. During the subsequent violent collision between G2 and G3 at $t\approx 1.2$ Gyr, further shocks and outflows are driven into the pre-shocked gas of the IGM (middle right panel in Fig. \ref{evolution_vrms_G6-IGM9}). The pre-shocked gas is now ejected by the cumulative shock fronts of this interaction, thus magnetizing the IGM. At $t\approx 1.8$ Gyr, the IGM within several 100 kpc around the collision debris is magnetized and highly turbulent (lower left panels in Figs. \ref{evolution_bfld_G6-IGM9} and \ref{evolution_vrms_G6-IGM9}). The debris continues to interact until their mutual final merger at $t\approx 2.5$ Gyr (lower right panels in Figs. \ref{evolution_bfld_G6-IGM9} - \ref{evolution_vrms_G6-IGM9}). By the end of the simulation, the galactic magnetic field has an average value of approximately $10^{-6}$ G, thus retaining its initial value, and the average IGM magnetic field reached a final strength of roughly $10^{-8}$ G (lower right panel in Fig. \ref{evolution_bfld_G6-IGM9}).

The global morphological evolution of the system within the G9-IGM9, G9-IGM12 and the G0-IGM0 scenarios is similar to the evolution within the G6-IGM9 scenario presented above, i.e. the evolution of the stellar distribution does not change significantly, and the collisions and merger events take place at the same times.\footnote{This is not surprising, as it is commonly believed that the global morphological evolution of interacting galactic systems is determined mainly by gravity and associated tidal action. However, it is interesting to note that in the early sixties, some authors believed that the filaments and tails observed in interacting galactic systems have nothing to do with tidal phenomena and even considered magnetic fields as the driving force for the morphological appearance of those systems (\citealp{Vorontsov-Velyaminov1960, Vorontsov-Velyaminov1962}).} For clarity, we do not show the corresponding figures to Figs. \ref{evolution_bfld_G6-IGM9} - \ref{evolution_vrms_G6-IGM9} for the other scenarios here (see appendix \ref{Appendix}), but rather highlight the main differences below.

\subsection{Differences between the scenarios}\label{Differences}

The most prominent differences between the different scenarios are the evolution of the magnetic field and, simultaneously, the behaviour of the shocks driven into the IGM during the interactions. Thereby, the shocks are propagating faster (and gain higher Mach numbers, section \ref{differences_mag}) for stronger initial magnetic fields. This is best visible during and shortly after the first collision ($t\approx 700-900$ Myr), when shocks are driven into the IGM for the first time. After the second collision, multiple shocks are propagating through the IGM which are additionally moving with different velocities depending on the scenario considered. Thus, even if it is possible to identify a single shock within one scenario, it is not possible to find the corresponding shock structure within the other scenarios. Therefore, within this section, we confine our analysis to the time before the second collision, where we can identify and compare the shocks within every scenario.

\subsubsection{Differences in the shock propagation}

\begin{figure}
\begin{center}
  \epsfig{file=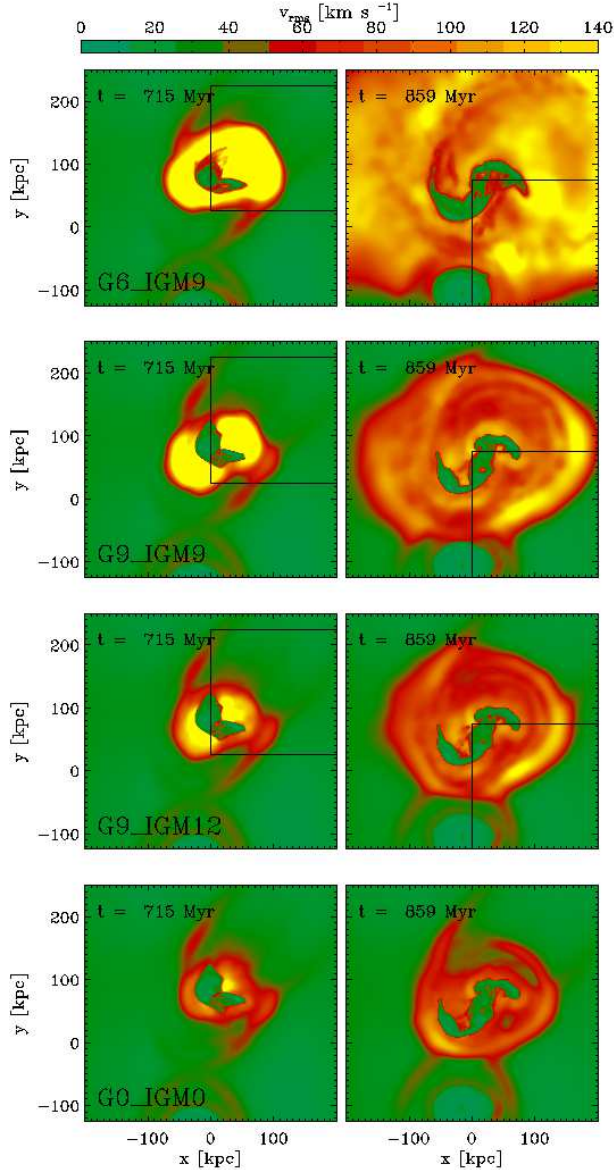, width=0.47\textwidth}
  \caption{Shock propagation within the different scenarios, traced by the rms velocity. From top to bottom: the mean line-of-sight rms velocity at $t=715$ Myr (left panels) and $t=859$ Myr (right panels) for the G6-IGM9, G9-IGM9, G9-IGM12 and the G0-IGM0 scenario, respectively. The color table is the same as in Fig. \ref{evolution_vrms_G6-IGM9}, whereby we only show the upper two galaxies G1 and G2. The higher the initial magnetic field (decreasing from top to bottom), the faster the shock propagation. The black boxes indicate the regions within which we analyse the shocks in more detail (see below).
  \label{shock_propagation}}
\end{center}
\end{figure}

Fig. \ref{shock_propagation} shows the shock propagation within the different scenarios, which is best visibly traced by the rms velocity. From top to bottom: the mean line-of-sight rms velocity at $t=715$ Myr (left panels) and $t=859$ Myr (right panels) for the G6-IGM9, G9-IGM9, G9-IGM12 and the G0-IGM0 scenario, respectively. The color table is the same as in Fig. \ref{evolution_vrms_G6-IGM9}, whereby we only show the upper two galaxies G1 and G2.

The energy initially released in form of shocks by the subsequent interactions should be comparable within each scenario, as the setup of the galaxies and thus the impact parameters are the same. Nevertheless, Fig. \ref{shock_propagation} clearly shows that the stronger the initial magnetic field (decreasing from top to bottom), the faster the shocks propagate (note that the different scenarios differ \textit{only} in the value of the initial field). The shock is the slowest for the scenario excluding magnetic fields (G0-IGM0, bottom panels). This behaviour suggests that the magnetic pressure associated with the magnetic field is able to additionally push the gas driven out by the interaction and thus accelerate the shocks.

\subsubsection{Differences in the magnetic field evolution}\label{differences_mag}

\begin{figure*}
\begin{center}
  \epsfig{file=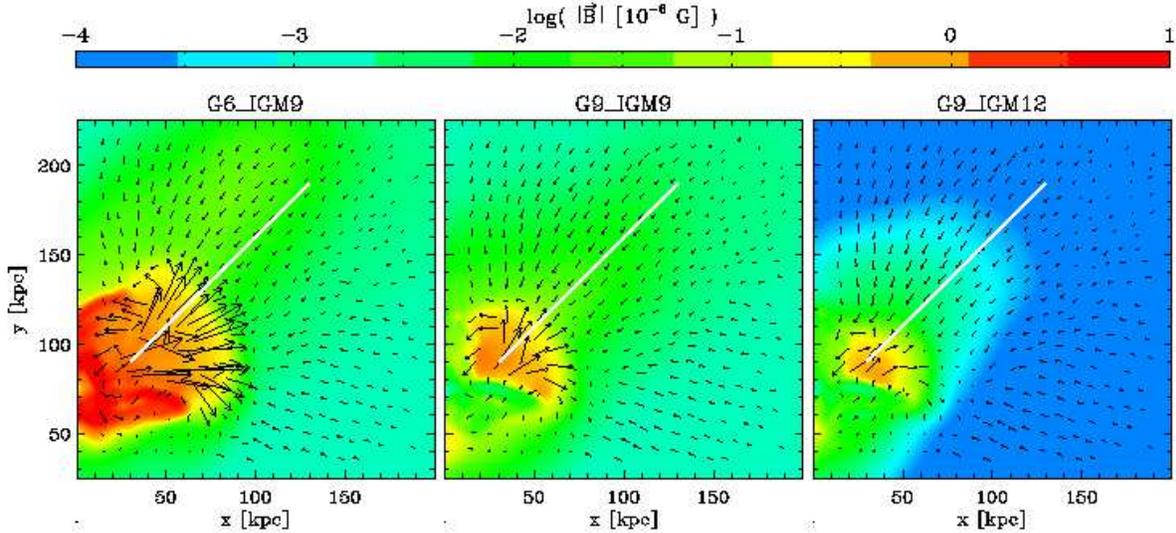, width=0.9\textwidth}
\end{center}
  \caption{Zoom-ins on the shock region at $t=715$ Myr (black boxes in the left panels in Fig. \ref{shock_propagation}) for the G6-IGM9 scenario (left), the G9-IGM9 scenario (central) and the G9-IGM12 scenario (right). Colours give the total magnetic field strength similar to Fig. \ref{evolution_bfld_G6-IGM9} and black vectors show the gas velocity (a length of ten on the given spatial scale correspond to 500 km s$^{-1}$). The white line indicates the line-of-sight $r_\mathrm{los}$ used for the shock analysis. The magnetic field strength reaches $\approx 10^{-6}$ G behind the shock within each scenario.
  \label{shock1b}}
\end{figure*}

Fig. \ref{shock1b} shows zoom-ins on the shock region at $t=715$ Myr (black boxes in the left panels in Fig. \ref{shock_propagation}) for the G6-IGM9 scenario (left panel), the G9-IGM9 scenario (central panel) and the G9-IGM12 scenario (right panel), respectively. Colours give the total magnetic field strength applying the same colour coding as in Fig. \ref{evolution_bfld_G6-IGM9}. Black vectors show the gas velocity whereby a length of ten on the given spatial scale corresponds to 500 km s$^{-1}$. The white line indicates the line-of-sight $r_\mathrm{los}$ used for the shock analysis presented below. The higher the initial magnetic field, the faster the shock (see also Fig. \ref{shock_propagation}), and the magnetic field is significantly strengthened behind the shock front. Independent of the initial magnetization model, the magnetic field strength behind the shock reaches values of the order of $10^{-6}$ G, thus showing that the growth is more efficient for lower initial magnetic fields.

\begin{figure*}
\begin{center}
  \epsfig{file=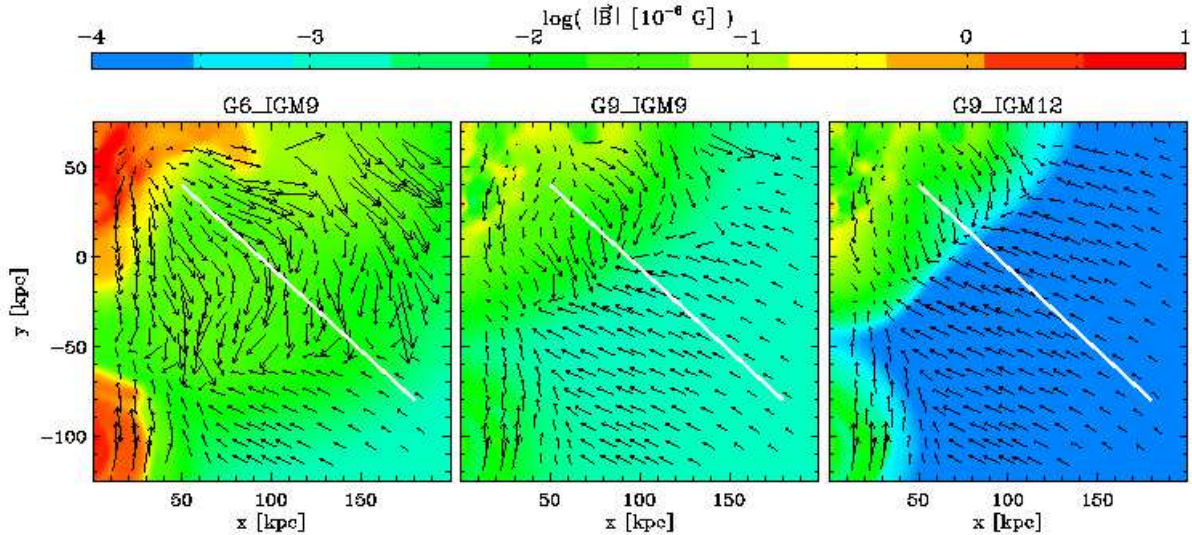, width=0.9\textwidth}
\end{center}
  \caption{Zoom-ins on the shock region at $t=859$ Myr (black boxes in the right panels in Fig. \ref{shock_propagation}), similar to Fig. \ref{shock1b}, but with a vector length of ten corresponding to 200 km s$^{-1}$. The IGM magnetic field strength behind the shock front reaches $\approx 10^{-8}$ G within each scenario.
  \label{shock2b}}
\end{figure*}

Fig. \ref{shock2b} shows zoom-ins on the shock at $t=859$ Myr (black boxes in the right panels in Fig. \ref{shock_propagation}), similar to Fig. \ref{shock1b}, but with vector length of ten corresponding to 200 km s$^{-1}$. As before, the magnetic field strengths behind the shock reach a common value. However, within the tenuous IGM gas, the magnetic field reaches strengths of only $10^{-8}$ G.

\begin{figure*}
\begin{center}
  \epsfig{file=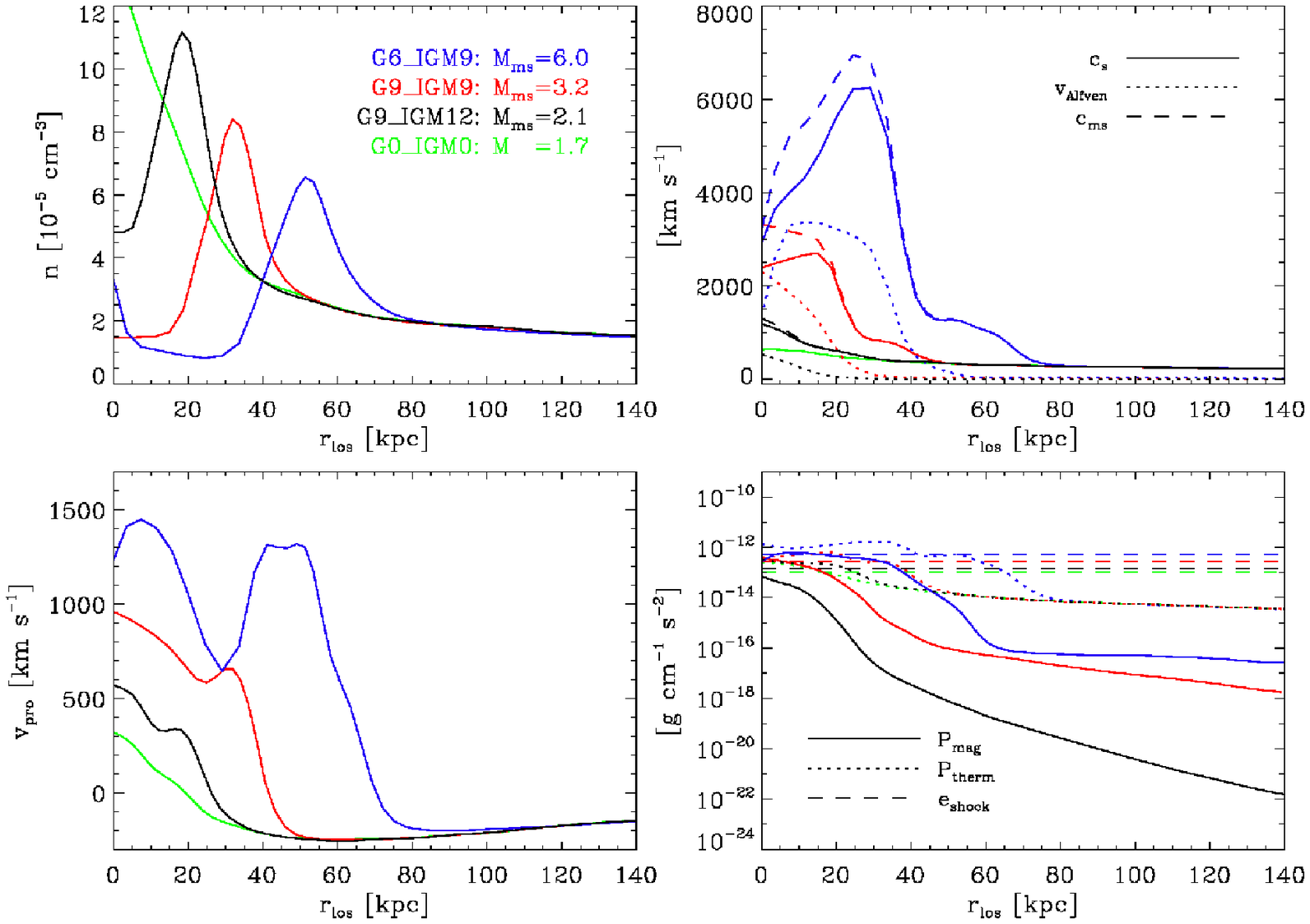, width=0.8\textwidth}
\end{center}
  \caption{Hydrodynamic values along the line-of-sight $r_\mathrm{los}$ at $t=715$ Myr for the G6-IGM9 scenario (blue lines), the G9-IGM9 scenario (red lines), the G9-IGM12 scenario (black lines), and the G0-IGM0 scenario (green lines), respectively. From left to right and top to bottom: The number density $n$; the sound velocity $c_\mathrm{s}$, the Alfven velocity $v_\mathrm{Alfven}$ and the magnetosonic velocity $c_\mathrm{ms}$, respectively; the gas velocity projected onto the line-of-sight $v_\mathrm{pro}$; and, eventually, the magnetic pressure $P_\mathrm{mag}$, the thermodynamical pressure $P_\mathrm{therm}$ and the shock energy density $e_\mathrm{shock}$, respectively. Magnetosonic Mach numbers $M_\mathrm{ms}$ for the magnetized scenarios are given in the upper left panel. For the G0-IGM0 scenario, we give the Mach number $M$ and do not show $v_\mathrm{Alfven}$, $c_\mathrm{ms}$ and $P_\mathrm{mag}$, respectively.
  \label{shock1}}
\end{figure*}

Fig. \ref{shock1} shows the hydrodynamic values along the line-of-sight $r_\mathrm{los}$ for the shock at $t=715$ Myr. The values are calculated using the SPH interpolation formalism. We show the shock properties for the G6-IGM9 scenario (blue lines), the G9-IGM9 scenario (red lines), the G9-IGM12 scenario (black lines), and the G0-IGM0 scenario (green lines), respectively. From left to right and top to bottom: The number density $n$; the sound velocity $c_\mathrm{s}$, the Alfven velocity $v_\mathrm{Alfven}=B/\sqrt{4\pi\rho}$ and the magnetosonic velocity $c_\mathrm{ms}=\sqrt{c_\mathrm{s}^2+v_\mathrm{Alfven}^2}$, respectively (whereby we do not show $v_\mathrm{Alfven}$ and $c_\mathrm{ms}$ for the G0-IGM0 scenario); the gas velocity projected onto the line-of-sight $v_\mathrm{pro}$; and, finally, the magnetic pressure $P_\mathrm{mag}$ (except for the G0-IGM0 scenario), the thermodynamical pressure $P_\mathrm{therm}$ and the shock energy density $e_\mathrm{shock}$, respectively. We estimate the shock energy density by considering the difference between the projected downstream (ds) peak velocity $v^\mathrm{ds}_\mathrm{pro,peak}$ (i.e. the shock velocity) and the average upstream (us) velocity $v^\mathrm{us}_\mathrm{pro}$, i.e. $e_\mathrm{shock}=1/2\rho^\mathrm{ds}\cdot(v^\mathrm{ds}_\mathrm{pro,peak}-v^\mathrm{us}_\mathrm{pro})^2$. Magnetosonic Mach numbers $M_\mathrm{ms}=(v^\mathrm{ds}_\mathrm{pro,peak}-v^\mathrm{us}_\mathrm{pro})/c_\mathrm{ms}^\mathrm{us}$ for the magnetized scenarios, and the Mach number $M=(v^\mathrm{ds}_\mathrm{pro,peak}-v^\mathrm{us}_\mathrm{pro})/c_\mathrm{s}^\mathrm{us}$ for the G0-IGM0 scenario are given in the upper left panel.

Again, Fig. \ref{shock1} clearly shows that the shock is propagating faster the stronger the initial magnetic field (e.g. lower left panel). Thus, it is the fastest within the G6-IGM9 scenario, and the slowest within the G0-IGM0 scenario. This increase in the shock intensity is also reflected in the projected gas velocities (lower left panel) and the corresponding Mach numbers, which range from 1.7 for the G0-IGM0 scenario to 6.0 for the G6-IGM9 scenario.\footnote{Mach numbers of approximately 2-5 have also been found by \citet{Johansson2009GravHeat} in their numerical studies of gravitational heating through the release of potential energy from infalling stellar clumps on galaxies.} Within each scenario (except for the G0-IGM0 scenario), the magnetic field gets strengthened behind the shock (solid lines in the lower right panel). Thereby, the magnetic pressure behind the shock is of order of the energy density of the shock itself (between $10^{-13}$ and $10^{-12}$ erg cm$^{-3}$) within each scenario, thus confirming that the growth of the magnetic field is the more efficient the lower the initial magnetic field.

\begin{figure*}
\begin{center}
  \epsfig{file=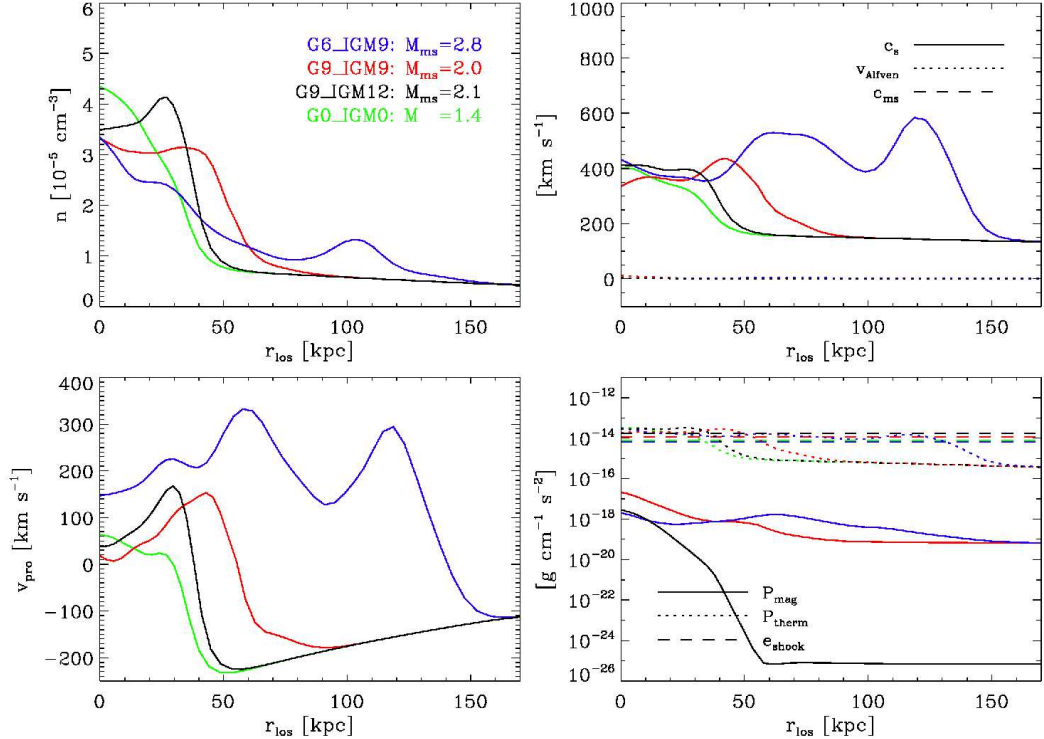, width=0.8\textwidth}
\end{center}
  \caption{Same as Fig. \ref{shock1}, but for the shock at $t=859$ Myr. Note the different $y$-ranges compared to Fig. \ref{shock1}.
  \label{shock2}}
\end{figure*}

Fig. \ref{shock2} shows the same quantities as in Fig. \ref{shock1} but for the shock at $t=859$ Myr. Again, the shock is propagating faster the stronger the initial magnetic field. However, the Mach numbers are lower than at $t=715$ Myr, ranging from 1.4 for the G0-IGM0 scenario to only 2.8 for the G6-IGM9 scenario. Thus, also the shock energies are lower by roughly one order of magnitude (lower left panel). As before, the magnetic field gets strengthened behind the shock (lower left panel). However, for this weaker subsequent shock which is propagating through a pre-shocked gas further outside the galaxies, the growth is not as efficient as at $t=715$ Myr. Hence, the magnetic pressure behind the shock does not reach the shock energy density.

\begin{figure*}
\begin{center}
  \epsfig{file=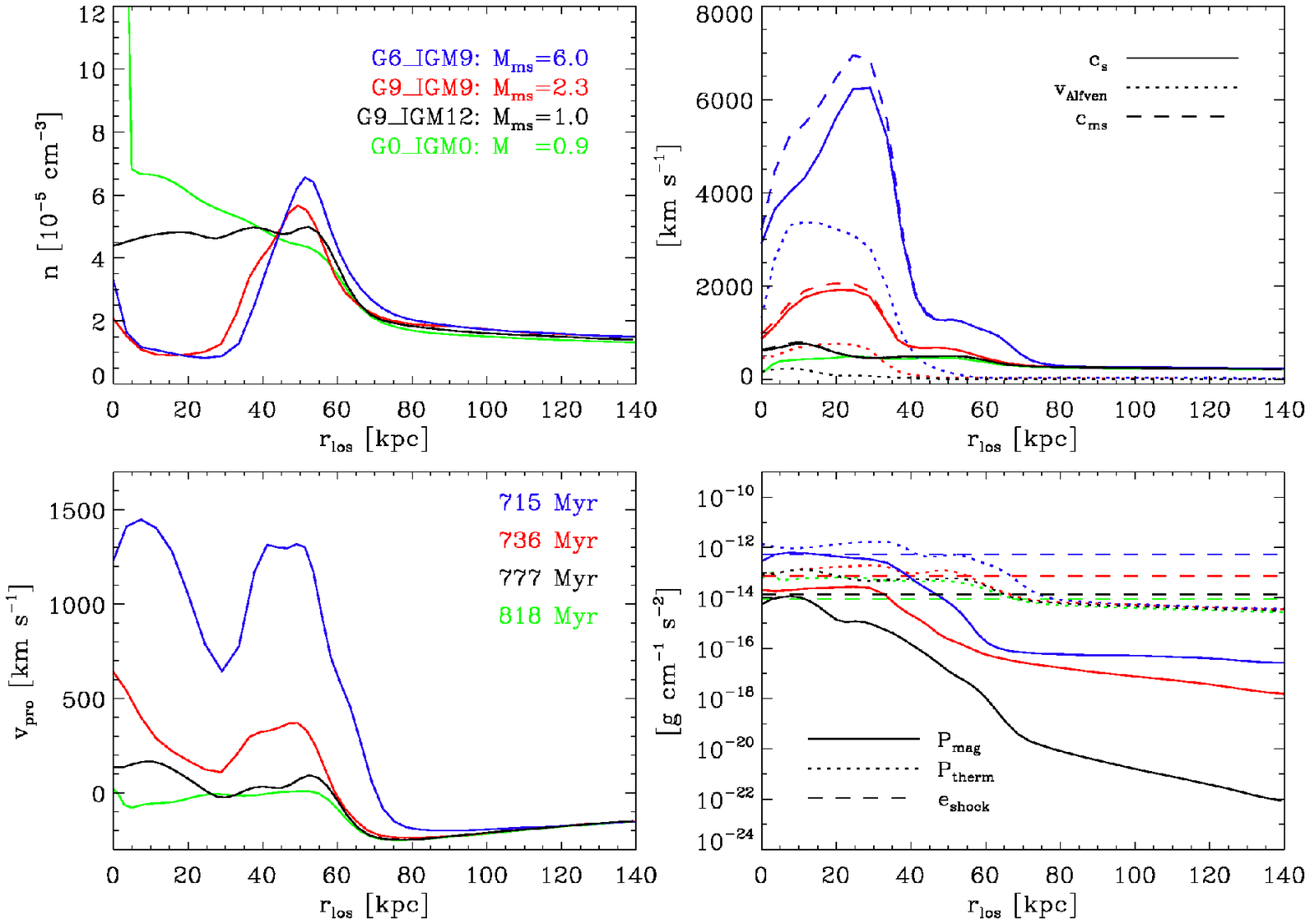, width=0.8\textwidth}
\end{center}
  \caption{Same as Fig. \ref{shock1}, but for these moments in time, at which the density peaks of the shocks driven by the first collision within each scenario have covered the same distance (the different times for each scenario are given in the lower left panel). The even greater difference in the Mach numbers compared to Fig. \ref{shock1} confirms the conclusion that the shocks are supported by magnetic pressure.
  \label{shock3}}
\end{figure*}

However, given the different sound speeds (or magnetosonic velocities) within the different scenarios, also the hydrodynamical timescales ($\sim c_\mathrm{s}/l$, where $l$ is a typical length-scale) are different. Thus, the comparison of the shocks at the same global dynamical time (defined by the gravitational interaction) might be questioned. In order to verify our conclusions drawn on the basis of Figs. \ref{shock1} and \ref{shock2}, we show how the shocks within each scenario compare when the density peaks of the shocks driven by the first collision have covered the same distance (Fig. \ref{shock3}). The corresponding different global dynamical times for each scenario are given in the lower left panel of Fig. \ref{shock3}. The lower the sound speed (upper right panel), the later in time the morphological agreement with the G6-IGM9 scenario (blue lines) is reached (upper left panel). The difference in the shock behaviour, i.e. the fact that the projected velocities are smaller for a smaller initial magnetization (lower left panel), is even more pronounced as before in Fig \ref{shock1}. This is reflected also in the Mach numbers (upper left panels in Fig. \ref{shock1} and \ref{shock3}) and in the energies of the shocks (lower right panels in Fig. \ref{shock1} and \ref{shock3}), and confirms our conclusion that the shocks are stronger for a stronger initial magnetization.

In summary, when shocks and outflows are driven by the interactions, the magnetic fields get strengthened. Thereby, the weaker the initial magnetic field, the more efficient the growth of the field (except for scenario G0-IGM0, where we exclude magnetic fields). The magnetic fields are thereby most efficiently strengthened behind the shocks. By the end of the simulations, the magnetic field strengths and distributions within every scenario are comparable (see section \ref{globalEvolutionMF} and the appendix \ref{Appendix_bfld}).

\subsubsection{Differences in the temperature evolution}

\begin{figure*}
\begin{center}
  \epsfig{file=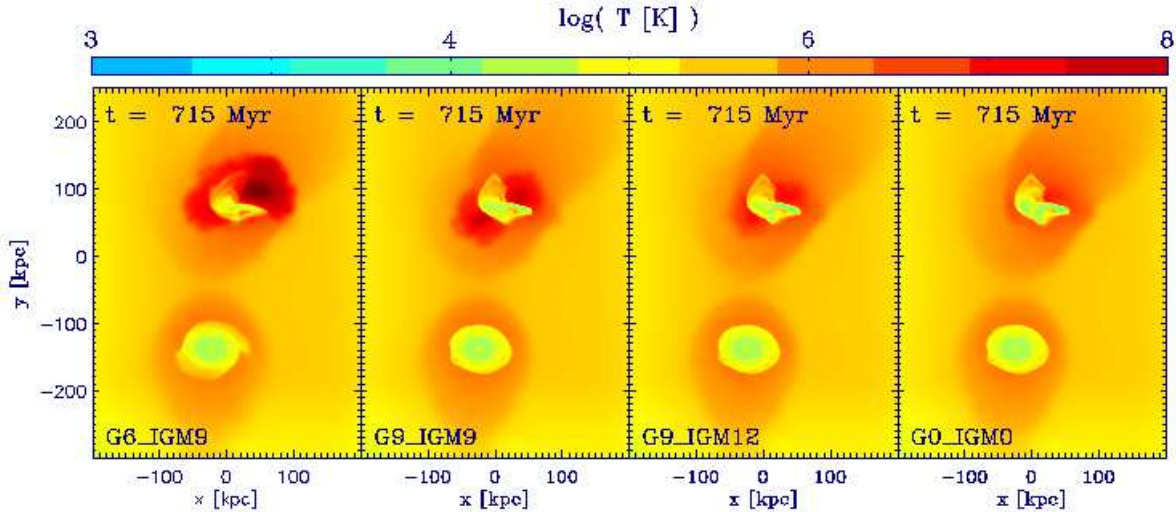, width=0.9\textwidth}
\end{center}
  \caption{Comparison of the temperatures at $t=715$ Myr within the different scenarios. From left to right: the G6-IGM9, G9-IGM9, G9-IGM12 and the G0-IGM0 scenario. The color table is the same as in Fig. \ref{evolution_temp_G6-IGM9}. The higher the Mach numbers (decreasing from left to right), the higher the temperature.
  \label{comparison_temp}}
\end{figure*}

According to the Rankine-Hugoniot shock jump conditions, the temperature behind the shock is proportional to the Mach number of the shock (more precisely, for $\gamma=5/3$, $T_\mathrm{ds}/T_\mathrm{us}\propto 5M^2+14-3/M^2$, see \citealp{LandauLifshitz1959}). Thus, the higher Mach numbers within the different scenarios are reflected in the temperatures of the IGM behind the shock fronts (best visible for $t=715$ Myr, Fig. \ref{comparison_temp}). Simultaneously, the region of shock heated gas is larger the stronger the shock (see also appendix \ref{Appendix_temp}). Thereby, the IGM temperatures reach values up to $10^8$ K. Gas of this temperature can be expected to give rise to significant X-ray emission by thermal bremsstrahlung. This was shown by e.g. \citet{Cox2006}, who performed hydrodynamical simulations of mergers of gas-rich disk galaxies. They showed that the hot diffuse gas that is produced by strong shocks attending the merger process can produce appreciable X-ray emission. Hence, it is interesting to note that the expected X-ray emission during galactic interactions increases with increasing magnetic field.\\

As the nature of the presented galactic interaction is highly non-linear, our conclusions should be taken as general trends. The behaviour of the shocks in our simulations strongly indicates that in the presence of a magnetic field the shocks driven by an interaction are supported by the magnetic pressure, thus resulting in higher Mach numbers. Simultaneously, the magnetic field evolution in our simulations is a strong indication for a shock driven magnetic field amplification. As most of the gas in the Universe is magnetized and shocks can be driven by different processes, we want to emphasize that the ``magnetic shock acceleration'' seen in our simulations might have implications within many astrophysical settings, e.g. galaxy clusters.

\subsection{Global evolution}\label{globalEvolution}

\subsubsection{Magnetic fields}\label{globalEvolutionMF}

\begin{figure}
\begin{center}
  \epsfig{file=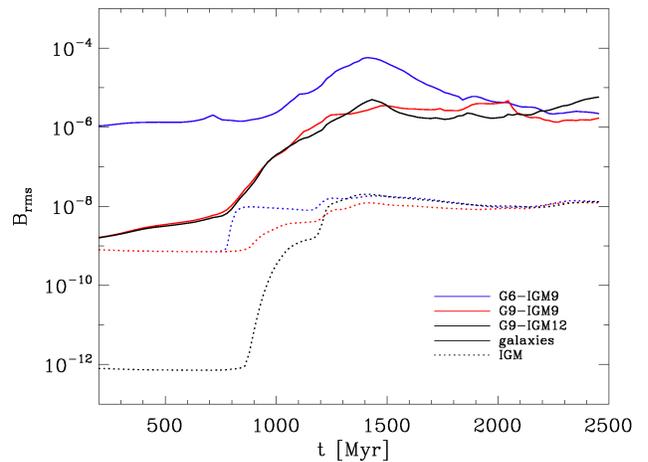, width=0.47\textwidth}
  \caption{$B_\mathrm{rms}=\sqrt{\langle B^2\rangle}$ as a function of time for the G6-IGM9 scenario (blue lines), the G9-IGM9 scenario (red lines), and the G9-IGM12 scenario (black lines). We separately show the IGM values (dotted lines) and the values inside the galaxies (solid lines). We distinguish between the IGM and the galaxies applying a density threshold of $10^{-29}$ g cm$^{-3}$. The final values for the galactic and the IGM magnetic field, respectively, are the same within all scenarios.
  \label{Bwitht}}
\end{center}
\end{figure}

Fig. \ref{Bwitht} shows the rms magnetic field $B_\mathrm{rms}=\sqrt{\langle B_x^2+B_y^2+B_z^2\rangle}$ as a function of time for the G6-IGM9 scenario (blue lines), the G9-IGM9 scenario (red lines), and the G9-IGM12 scenario (black lines). We separately plot the IGM values (dotted lines) and the values inside the galaxies (solid lines). We distinguish between the IGM and the galaxies applying a density threshold of $10^{-29}$ g cm$^{-3}$.

Within all magnetized scenarios, the  galactic magnetic field saturates at several $\mu$G by the end of the simulation, independent of the initial magnetic field strength. The first collision between G1 and G2 initiates within all scenarios a steady growth of the galactic rms magnetic field, which continues throughout the second and third collision. This growth starts at a later time and is less efficient within the G6-IGM9 scenario (where the initial galactic field is already of order of the final value) than within the other scenarios. This indicates that it is easier to enhance a magnetic field which is not yet of the order of the saturation value. In the subsequent evolution, the galactic rms magnetic field strength within the G6-IGM9 scenario decreases again. Thus, at time of the intermediate merger ($t\approx 1.8$ Gyr), the galactic magnetic field has a rms value between 1 and 10 $\mu$G within all scenarios. The intermediate and the final mergers ($t\approx 2$ Gyr) do not lead to further growth of the field.

The evolution of the IGM rms magnetic field (dotted lines) results in a common final value of approximately $10^{-8}$ G within all scenarios. Again, the first collision initiates a growth of the IGM magnetic field. Within the G9-IGM12 scenario, where the initial IGM field is four orders of magnitude smaller than the final value, this growth continues throughout the second and third collision until the final value is reached after approximately 500 Myrs. Within the G6-IGM9 scenario, the initial IGM field of $10^{-9}$ G grows by a factor of 10 at time of the first collision, thus reaching the final value shortly after this collision. None of the further collision and merger events lead to a significant further growth of the field. Most interestingly, the same initial IGM field of $10^{-9}$ G does not grow as efficiently at time of the first collision within the G9-IGM9 scenario. Rather, within the G9-IGM9 scenario, the IGM field grows more slowly and reaches the final value only after the final merger. The only difference between the G9-IGM9 and G6-IGM9 scenarios is the different initial galactic magnetic field, which is three order of magnitude higher within the G6-IGM0 scenario. Thus, the faster growth of the IGM field within the G6-IGM9 scenario shows that interaction-driven outflows transport magnetic field energy from the galaxies into the IGM, resulting in a more efficient growth of the IGM magnetic field.

The fundamental source of energy for the general strengthening of the magnetic field is the gravitational energy released during the interaction of the galaxies. This gravitational energy is converted into kinetic energy of the particles, particularly, turbulence which is expected to be driven on kpc scales. The kinetic energy in turn is partly converted into magnetic energy by compression, shearing and folding of the magnetic field lines. This behaviour results from the basic thermodynamical principle according to which the free energy of a system will always be distributed among all available energy channels. Given the high variability of the simulated system and the limited resolution, however, detailed studies of the small-scale processes responsible for this distribution (e.g. the turbulent dynamo) are beyond the scope of this paper. Yet, simple magnetic field line compression would result in a tight correlation between the magnetic field strength and the gas density. Thus, it is possible to demonstrate the presence of the other processes, i.e. the shearing and folding of the magnetic field lines in the course of turbulent amplification, by analyzing the distribution of the magnetic field strength as a function of the gas density.

\begin{figure*}
\begin{center}
  \epsfig{file=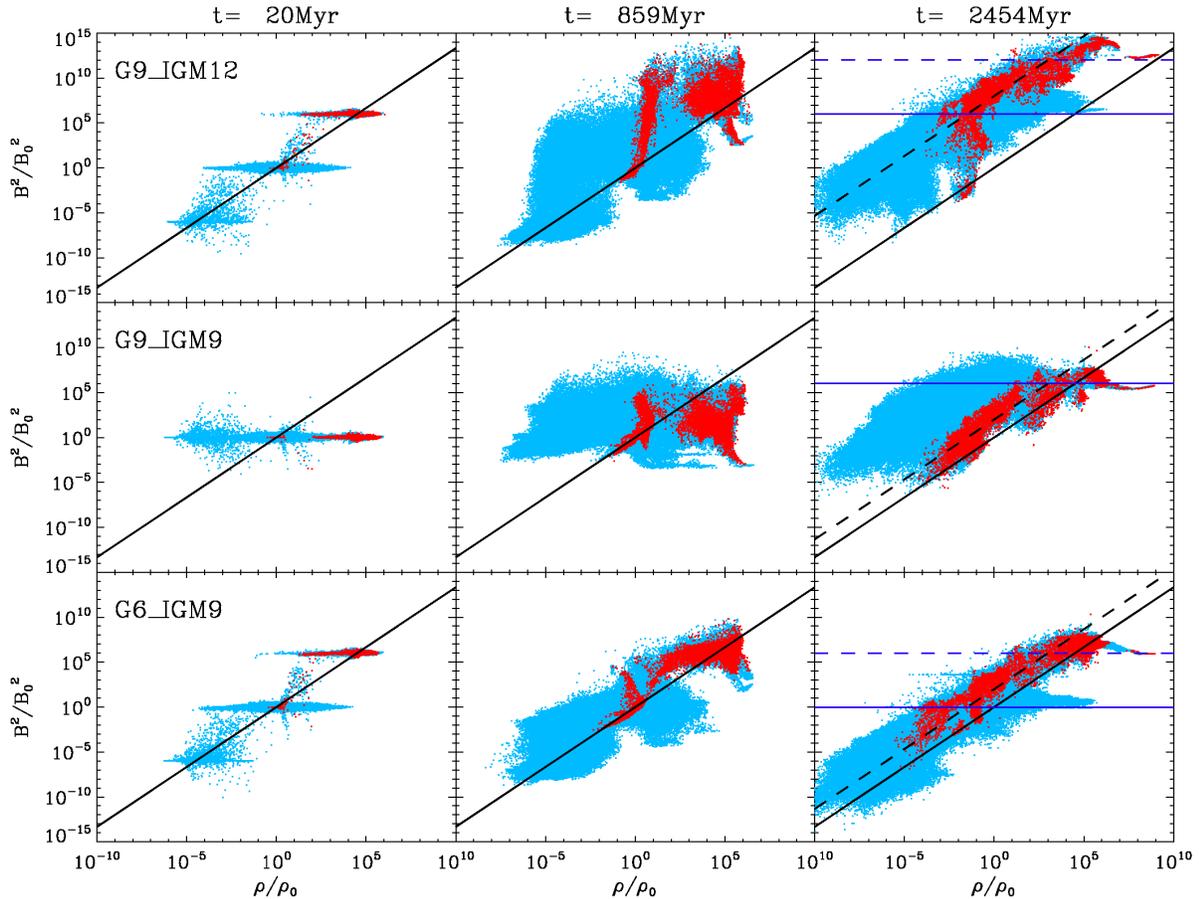, width=0.9\textwidth}
\end{center}
  \caption{$B^2/B_0^2$ as a function of $\rho/\rho_0$ at three different timesteps (from left to right: $t=20$~Myr, $t=859$~Myr and $t=2454$~Myr) for the G9-IGM12 scenario (upper panels), the G9-IGM9 scenario (central panels) and the G6-IGM9 scenario (lower panels). The light blue dots correspond to particles  with an initial density $\rho \geq 10^{-29}$~g~cm$^{-3}$ (i.e. galactic particles), and the residual particles (IGM) are marked in red. The black solid lines follow the relation $B^2/B_0^2=(\rho/\rho_0)^{4/3}$, which is valid for isotropic compression. The final distributions of the magnetic field strengths (right panels) lie above this relation, showing that compression is not the only process responsible for the growth of the magnetic field. The black dashed lines (right panels) show the same relation but shifted by the factor by which the IGM magnetic field has been strengthened within each scenario (not fitted to the distribution). The horizontal solid and dashed dark blue lines (right panels) correspond to the difference between, respectively, the initial galactic and IGM magnetic field ($B_\mathrm{0,disk}$ and $B_\mathrm{0,IGM}$) within each scenario, and the common final galactic rms value of $10^{-6}$~G (for the G9-IGM9 scenario, these lines lie on top of each other). These are the saturation levels of the magnetic field in the different scenarios.
  \label{scatter}}
\end{figure*}

Fig. \ref{scatter} shows $B^2/B_0^2$ as a function of $\rho/\rho_0$ at three different timesteps (from left to right: $t=20$~Myr, $t=859$~Myr and $t=2454$~Myr) for the G9-IGM12 scenario (upper panels), the G9-IGM9 scenario (central panels) and the G6-IGM9 scenario (lower panels). The light blue dots correspond to particles  with an initial density $\rho \geq 10^{-29}$~g~cm$^{-3}$ (i.e. galactic particles), and the residual particles (IGM) are marked in red. The ``trimodality'' of the distribution of the G9-IGM12 and the G6-IGM9 scenarios at $t=20$~Myr is a result of the initial conditions: At the edges of the galaxies, IGM particles carrying a weak magnetic field overlap with galactic particles having a stronger field. Magnetic diffusion redistributes the field among these edge particles, leading to an increase of the IGM field and a decrease of the galactic field in these regions. This effect is of course not seen for the G9-IGM9 scenario, where all particles carry the same initial magnetic field.

The black solid lines crossing all panels of Fig. \ref{scatter} follow the equation $B^2/B_0^2=(\rho/\rho_0)^{4/3}$, which is valid for isotropic compression of a turbulent magnetic field. If compression would be the only process responsible for the magnetic field growth, the distribution could evolve only along this line. Yet, the intermediate and final distributions of the magnetic field strengths (middle and right panels) lie above this relation, showing that turbulent amplification has additionally strengthened the field (this is least prominent for the G6-IGM9 scenario, where the initial magnetic field strength is already of the order of the final strength). On the other hand, the final distributions (right panels) follow the $4/3$ inclination, thus showing that compression has to be a part of the overall strengthening process. The factor by which the IGM magnetic field (squared) gets strengthened is approximately $(10^{4})^2$ for the G9-IGM12 scenario and $(10)^2$ for the G9-IGM9 and the G6-IGM9 scenarios, respectively. The black dashed lines (right panels) correspond to the black solid lines but shifted by this factor (not fitted to the distribution), thus additionally demonstrating the importance of non-compressive processes for the magnetic field evolution.

The horizontal solid and dashed dark blue lines in Fig. \ref{scatter} (right panels) indicate the difference between, respectively, the initial galactic and IGM magnetic field ($B_\mathrm{0,disk}$ and $B_\mathrm{0,IGM}$) within each scenario, and the common final galactic rms value of $10^{-6}$~G (for the G9-IGM9 scenario, these lines lie on top of each other). As this final galactic rms value is the maximal magnetic field strength which can be acquired by both, the galactic and the IGM particles, the dark blue lines actually show the saturation levels of the magnetic field in the different scenarios. This is also visible from the ``noses'' on the right-hand side of the final distributions, which show that particles which have already reached the saturation level can not acquire higher magnetic field strengths even if the density increases.

In summary, Fig. \ref{scatter} shows that the magnetic field growth driven by the interactions is due to both, the compression and the folding and stretching of the magnetic field lines in turbulent flows. It also confirms the saturation level of the magnetic field strength as shown in Fig. \ref{Bwitht}. However, we note that the presented simulations are not designed to study the process of turbulent amplification in detail, wherefore our results should be interpreted in terms of general thermodynamics and magnetohydrodynamics.

\subsubsection{Numerical divergence}\label{globalEvolutionDIVB}

\begin{figure}
\begin{center}
  \epsfig{file=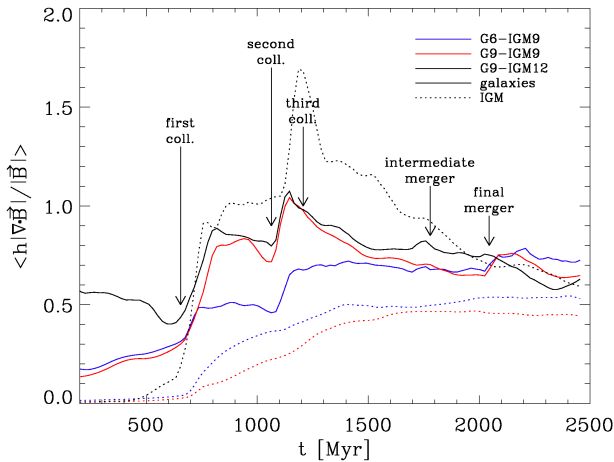, width=0.47\textwidth}
  \caption{The mean $\langle h\nabla\cdot\mathbf{B}/|\mathbf{B}|\rangle$ as a function of time for the G6-IGM9 scenario (blue lines), the G9-IGM9 scenario (red lines), and the G9-IGM12 scenario (black lines). We separately show the IGM values (dotted lines) and the values inside the galaxies (solid lines). We distinguish between the IGM and the galaxies applying a density threshold of $10^{-29}$ g cm$^{-3}$. The numerical divergence is small enough to guarantee numerical stability.
  \label{divBwitht}}
\end{center}
\end{figure}

Fig. \ref{divBwitht} shows the mean numerical divergence measure $\langle h_\mathrm{SPH}|\nabla\cdot \mathbf{B}|/|\mathbf{B}|\rangle$ as a function of time for the G6-IGM9 scenario (blue lines), the G9-IGM9 scenario (red lines), and the G9-IGM12 scenario (black lines). We separately show the IGM values (dotted lines) and the values inside the galaxies (solid lines). A numerical divergence is expected to arise from the SPH divergence operator even for a field which is divergence free in the first place, but tangled on sub-resolution scales. This numerical divergence arises also in simulations where the magnetic field is expressed in terms of Euler potentials, which avoid physical divergence by definition. Comparisons of simulations using the Euler potentials with simulations using the direct implementation (as in this paper) have shown that the numerical divergence does not influence the evolution of the magnetic field significantly as long as the divergence measure is $\leq 1$ (\citealp{me2009}). For the simulations presented here, the mean numerical divergence of all gas particles stays always below this tolerance value, except for the IGM value within the G9-IGM12 scenario during the third collision. Note, however, that this comparatively high divergence does not lead to an enhanced magnetic field growth (Fig. \ref{Bwitht}). Generally, the higher the initial magnetic field, the lower the numerical divergence. This trend is consistent with previous studies (\citealp{meAnt2009}). The reason for this behaviour is the Lorentz force acting on the particles, which is always opposing the motions leading to a change (growth) of the magnetic field. The stronger the field, the stronger this force. Thus, in the presence of a stronger field, random particle motions on sub-resolution scales are more efficiently suppressed. These motions result in a sub-resolution tangling of the magnetic field, and by reducing the motions, the numerical divergence is simultaneously reduced. Altogether, the numerical divergence should not influence the general conclusions on the magnetic field evolution presented in this paper (see also \citealp{me2009,meAnt2009}).

\subsubsection{Pressures}\label{Pressures}

Saturation phenomena as seen for the magnetic fields in our simulations (Fig. \ref{Bwitht}) usually suggest some kind of energy equipartition. Given enough time, a thermodynamical system will always distribute its free energy to all the degrees of freedom available. A particular saturation value is thereby reached when the energy which is increasing balances the energy of the source responsible for the increase. For example, in case of the galactic dynamo, the main source for the amplification of the magnetic field is the energy of the particles which have a velocity component perpendicular to the galactic disc. The corresponding rising and descending particle motions are expected to become helical under the influence of the Coriolis force, resulting in the so-called $\alpha$-effect (see e.g. \citet{Kulsrud1999} for a review). Assuming these particles to be cosmic ray (CR) particles rises the expectation of equipartition between the magnetic and the CR pressure (\citealp{Hanasz_etal2009}). This equipartition was recently shown by \citet{Hanasz_etal2009b}.
More generally, in the framework of MHD, any motion of the gas leading to a growth of the magnetic field will be suppressed by the magnetic field itself via the Lorentz force as soon as the magnetic energy gets comparable to the kinetic energy of the gas. The magnetic energy is then converted into kinetic energy of the gas, thus maintaining equipartition between the magnetic and gas kinetic energy. In particular, the magnetic field is expected to be in equipartition with the turbulent energy of the gas, as only velocity gradients can lead to a growth of the magnetic field. This concept of a ``turbulent dynamo'' is well known from theory (see e.g. \citealp{BrandenburgSubramanian2005} for a review). Thus, within this section, we analyse our simulations with respect to the different pressure components within the galaxies and within the IGM. \\

\begin{figure}
\begin{center}
  \epsfig{file=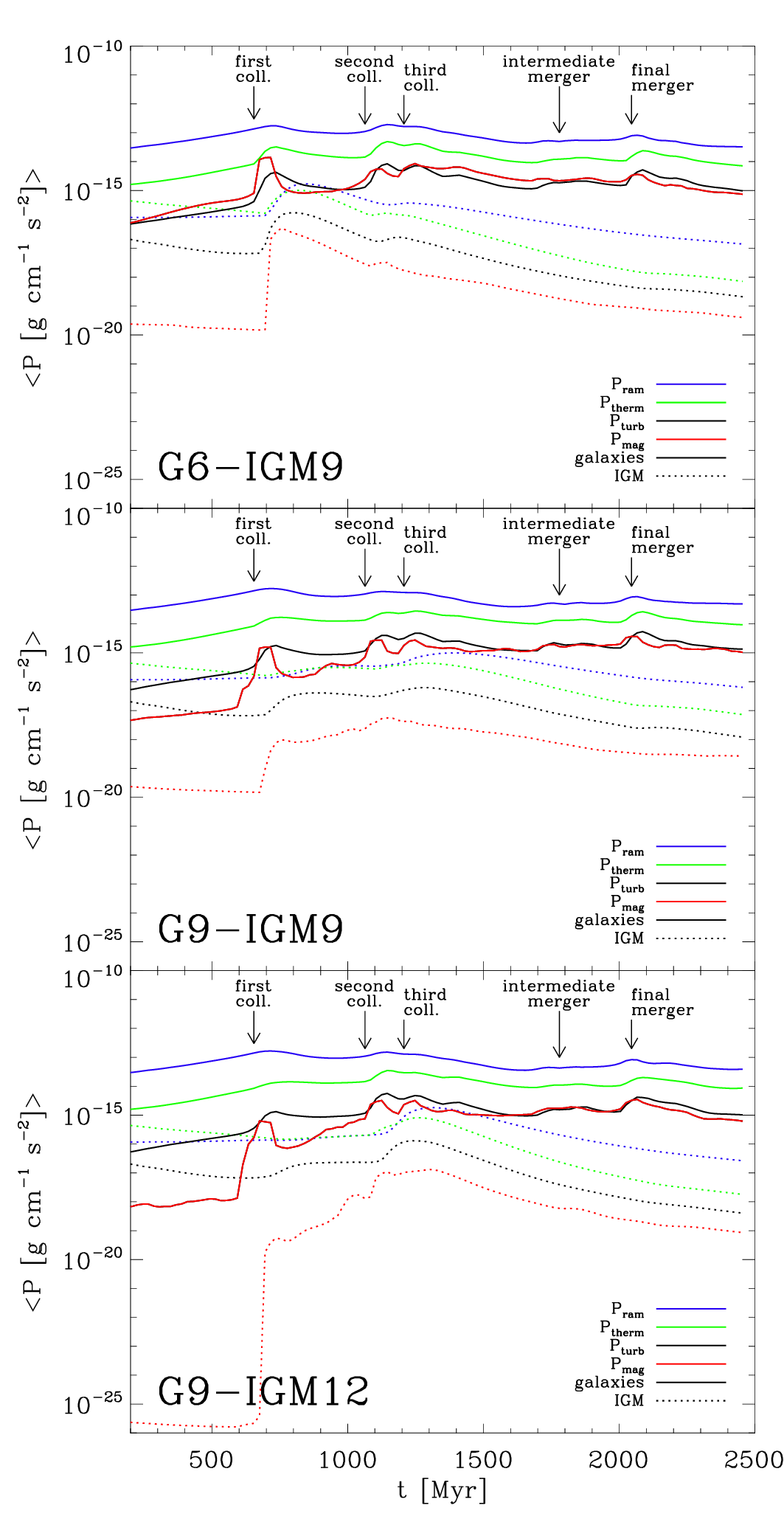, width=0.47\textwidth}
  \caption{Evolution of the volume weighted mean of different pressure components as a function of time for the G6-IGM9 scenario (upper panel), the G9-IGM9 scenario (central panel) and the G9-IGM12 scenario (lower panel). We plot the following pressures: magnetic pressure $P_\mathrm{mag}$ (red lines), thermal pressure $P_\mathrm{therm}$ (green lines), turbulent pressure $P_\mathrm{turb}$ (black lines) and ram pressure $ P_\mathrm{ram}$ (blue lines). We distinguish between the IGM (dotted lines) and the galaxies (solid lines) applying a density threshold of $10^{-29}$ g cm$^{-3}$. By the end of the simulations, the magnetic pressure is of the same order of magnitude as the turbulent pressure within all scenarios.
  \label{Pres_t}}
\end{center}
\end{figure}

Fig. \ref{Pres_t} shows the evolution of the volume weighted mean of different pressure components as a function of time for  the G6-IGM9 scenario (upper panel), the G9-IGM9 scenario (central panel) and the G9-IGM12 scenario (lower panel). We plot the following pressures: magnetic pressure $P_\mathrm{mag}=B^2/8\pi$ (red lines), thermal pressure $P_\mathrm{therm}=1/2\rho c_s^2$ (green lines, with $c_s$ being the sound speed and $\rho$ the gas density), turbulent pressure $P_\mathrm{turb}=1/2\rho v_\mathrm{rms}^2$ (black lines) and ram pressure $P_\mathrm{ram}=1/2\rho v^2$ (blue lines, with $v$ the total velocity of the gas particle considered, i.e. including turbulent components). Again, we distinguish between the IGM (dotted lines) and the galaxies (solid lines) applying a density threshold of $10^{-29}$ g cm$^{-3}$.

Within the galaxies (solid lines), the thermal and ram pressures within each scenario evolve smoothly with time, whereby the thermal pressure stays always below the ram pressure by roughly one order of magnitude. By the end of the simulations, these pressure components have the same values as at the beginning. The turbulent pressure (black lines) within each scenario increases by slightly more than one order of magnitude during the simulations, whereby each collision and merger event tends to increase the turbulent pressure.

The galactic magnetic pressures (red solid lines) within each scenario evolve in a similar way as the turbulent pressures, increasing at the times of collision and merger events and partly decreasing afterwards (due to the dilatation of the shocked region and the corresponding dilution of the magnetic field energy). After the second collision, the magnetic pressure is in approximate equipartition with the turbulent pressure within all scenarios. Before, the evolution of the magnetic pressures within the different scenarios differs because of the different initial magnetization. Within the G6-IGM9 scenario (upper panel), where the initial magnetic pressure is already of the order of the turbulent pressure, it increases by roughly one order of magnitude during the first collision, thus even exceeding the turbulent pressure. However, it decreases again to its initial value shortly after the first collision. Within the other scenarios, where the initial magnetic pressure is lower than the turbulent pressure, the galactic magnetic pressure increases during the first collision only up to the equipartition level, and does not decrease to its initial value after this collision.

Within the IGM (dotted lines), the thermal and ram pressures within each scenario are near equipartition until the third collision. This is reasonable, as we assume the IGM temperature to be the same as the virial temperature of the haloes at the beginning of the simulations. However, at roughly the time of the third collision, the thermal pressure begins to decrease more efficiently than the ram pressure, and, in the subsequent evolution, stays below the ram pressure by a factor 10 - 30. This bahaviour is similar for each scenario. The IGM turbulent pressure (black dotted lines) differs form the IGM ram pressure by roughly two orders of magnitude during the whole simulation within each scenario, which is comparable to the ratios between the turbulent and ram pressures within the galaxies.

The evolution of the IGM magnetic pressure (red dotted lines) follows the evolution of the turbulent pressure within each scenario after the second collision. This parallel evolution indicates a correlation between these pressure components. However, contrary to what is seen in the galaxies (solid lines), the IGM magnetic pressure stays by roughly one order of magnitude below the IGM turbulent pressure until the end of the simulations. This difference is most probably a result of the numerical method by which we estimate the turbulence, within which shearing and inhomogeneous motions can lead to an overestimate of the turbulence. This can particularly happen within the IGM, where shocks propagate in a nongeneric direction. Thus, the similarity of the evolution of the IGM magnetic and turbulent pressures may be interpreted as equipartition between these components.

Before the second collision, the evolution of the IGM magnetic pressure within the different scenarios again differs because of the different initial magnetization. Within the G6-IGM9 scenario (upper panel), the ``quasi'' equipartition with the turbulent pressure is reached already after the first collision, whereas it is reached only after the second collision within the other scenarios. Particularly, the IGM magnetic pressure within the G9-IGM9 (central panel) scenario does not grow as efficiently as within the G6-IGM9 scenario, although the initial IGM magnetization is the same within both scenarios. This difference is due to the stronger initial galactic magnetic field within the G6-IGM9 scenario: The strong galactic field is transported by interaction-driven outflows into the IGM, thus enhancing the IGM magnetic field more efficiently than within the G9-IGM9 scenario (see also section \ref{globalEvolutionMF}).\\

Summing up, we conclude that the saturation value within the galaxies of several $\mu$G seen in all simulations corresponds to the equipartition between turbulent and magnetic pressure, a result which was also achieved by \citet{meAnt2009}. Also, the saturation value within the IGM ($\approx 10^{-8}$ G) supports the assumption of equipartition between magnetic and turbulent pressure.

\section{Synthetic radio emission, polarization and $RM$ maps}\label{RADIO}

Radio observations of local galaxies and galaxy groups provide maps of the distribution and structure of the magnetic fields in these systems. However, the radio maps are always only one snapshot in time and observers face the problem of explaining the origin of the observed magnetic field strengths and structures. For example, observations of the compact group of galaxies Stephan's Quintet, which consists of four interacting spiral galaxies and was recently modeled by means of \textsl{N}-body simulations by \citet{Renaud2010}, show a prominent ridge of radio emission crossing through the system in between the galaxies (\citealp{Xu2003}). It is commonly believed that this ridge of radio emission corresponds to a shock front driven by a former interaction between two of the galaxies of the group. Due to the shock compression, the magnetic field of the ambient gas might have been amplified. Moreover, electrons are expected to get Fermi-accelerated within the shock, thus giving rise to the enhanced radio emission.

It is the aim of this section to show how our simulated system of three merging galaxies might look like when observed at a typical radio frequency. As we are able to provide synthetic radio maps at every snapshot, we can assess the evolution and thus the origin of the magnetic field pattern seen in the synthetic maps. As detailed radio observations are available only for local groups of galaxies, we have chosen the G6-IGM9 scenario for the calculation of the synthetic maps.

\citet{meAnt2009} have already presented such synthetic radio maps for an isolated galaxy and the Antennae system. Compared to their calculations, in this paper, we use a more precise formula for the synchrotron emission, which, however, does not change the results qualitatively (see e.g. \citealp{Ginzburg&Syrovatskii1965}, \citealp{Rybicki&Lightman1986} or \citealp{Longair1994} for more details on the standard theory of synchrotron radiation). Assuming an energy distribution of the relativistic CR electrons of a power-law form
\begin{equation}
  n(E)dE=\kappa E^{-p}dE\label{powerSpectrum}
\end{equation}
(where $n(E)$ is the number density of electrons, $\kappa$ a constant normalization factor and $p$ the index of the power spectrum), the synchrotron emissivity $J_\nu$ at a given frequency $\nu$ is given by
\begin{eqnarray}
  J_\nu &=& \frac{\sqrt{3}e^2 B_\bot \kappa}{m_e c^2 (p+1)}\left(\frac{m_e^3 c^5 2\pi\nu}{3eB_\bot}\right)^{-\frac{p-1}{2}}\notag\\
   & & \times\Gamma\left(\frac{p}{4}+\frac{19}{12}\right)\Gamma\left(\frac{p}{4}-\frac{1}{12}\right),
\end{eqnarray}
where $\Gamma$ denotes the Gamma function. The magnetic field component perpendicular to the line of sight $B_\bot$ is taken from the simulation. The frequency $\nu$ and the index $p$ are input parameters. Similar to \citet{meAnt2009} we assume $p=2.6$ and $\nu=4.86\times10^9$ Hz, corresponding also to the values given by \citet{Xu2003}. The value of $p$ gives a spectral index of the radio emission of $\alpha=(p-1)/2=0.8$, which is typical for spiral galaxies (\citealp{Gioia1982}), and given also by \citet{Xu2003} and \citet{ChyzyBeck2004}. Also, this value is close to the power-law slope of 2.7 of the all-particle spectrum of CRs in the energy rage of $10\times10^{9}\mbox{~eV}\leq E \leq 3\times10^{15}\mbox{~eV}$ (e.g. \citealp{Blasi2008} and references therein). The constants are the speed of light $c$, the electron mass $m_e$ and the electron charge $e$. For a given CR energy density $e_\mathrm{CR}$, $\kappa$ can be derived using Eqs. \ref{powerSpectrum}:
\begin{equation}
  e_\mathrm{CR}=\int_{E_\mathrm{min}}^{E_\mathrm{max}}En(E)dE=\kappa\int_{E_\mathrm{min}}^{E_\mathrm{max}}E^{1-p}dE.
\end{equation}
Similar to \citet{meAnt2009} we assume an energy range of $E_\mathrm{min}=10^9$ eV to $E_\mathrm{max}=10^{15}$ eV in our calculations. The cutoff at $10^9$~eV is justified, since for a magnetic field strength of 10 $\mu$G (which is the typical maximum field strength in our simulations) synchrotron emission  is radiated mainly by particles with an energy of $\approx 7\times 10^9$~eV (\citealp{Longair1994}, eq. 18.51). For lower magnetic fields, the energy of the electrons responsible for the observed radiation is even higher.

CRs with energies between $10^9$~eV and $10^{15}$~eV are produced mainly through particle acceleration within the shocks of SN remnants. They propagate by advection and diffusion along the magnetic field lines throughout the galaxy or the IGM. Given the non-triviality of the physics involved in these processes (e.g. \citealp{Blasi2008}), the distribution of CRs within a galactic system is not easily modeled. However, CRs are bound to the magnetic field, which, in turn, is coupled to the (ionized) thermal gas. Thus, as a reasonable but simple assumption, the CR energy distribution may be expected to be proportional to the thermal energy distribution. Particularly, we assume that $e_\mathrm{CR}=0.01\cdot e_\mathrm{therm}$.

$J_\nu$ is calculated for every particle, and the total synchrotron intensity $I_\mathrm{tot}$ is subsequently obtained by integrating along the line-of-sight. The polarized emission $I_\mathrm{pol}$, the predicted observed degree of polarization $\Pi_\mathrm{obs}$ and the polarization angles $\psi$ are calculated in the same way as in \citet{meAnt2009}.

The calculations are implemented within the code \textsc{P-Smac2} (Donnert et al., in preparation), which subsequently bins the values of $I_\mathrm{tot}$, $I_\mathrm{pol}$, $\Pi_\mathrm{obs}$ and $\psi$ on a grid using the gather approximation as before for Figs. \ref{evolution_bfld_G6-IGM9} to \ref{evolution_vrms_G6-IGM9}. In order to account for the spatial isotropy of the emission, we multiply the calculated total and polarized intensities by the factor
\begin{equation}
  f_\mathrm{obs}=\frac{d_\mathrm{pix}^2}{4\pi\cdot d^2},
\end{equation}
with $d$ the assumed distance to the observer and $d_\mathrm{pix}$ the pixel size (corresponding to a beam diameter) of $\approx 0.98$ kpc/$h$. We assume $d=80$ Mpc, which is a typical distance for local groups, e.g. the Stephan's Quintet (\citealp{Xu2003}). Thus, $d_\mathrm{pix}$ corresponds to an angular resolution of $\approx1''.3$. For comparison, the maximum resolution of the Very Large Array (VLA) at $\nu=4.86\times10^9$ Hz ($\lambda\approx6$~cm) is $0''.4$, and for the Effelsberg telescope $2''.4$. However, observations performed with these telescopes are usually presented with a lower resolution of $\approx 5''-15''$, e.g. the Stephan's Quintet (\citealp{Xu2003}), the Antennae galaxies (\citealp{ChyzyBeck2004}) or M51 (\citealp{Beck2009}). The upcoming Square Kilometer Array (SKA) is expected have 50 times the sensitivity of the VLA with a maximum angular resolution of $0''.02$ at $1.4$ GHz ($\lambda\approx21$~cm) (\citealp{Gaensler2009}).

\begin{figure*}
\begin{center}
  \epsfig{file=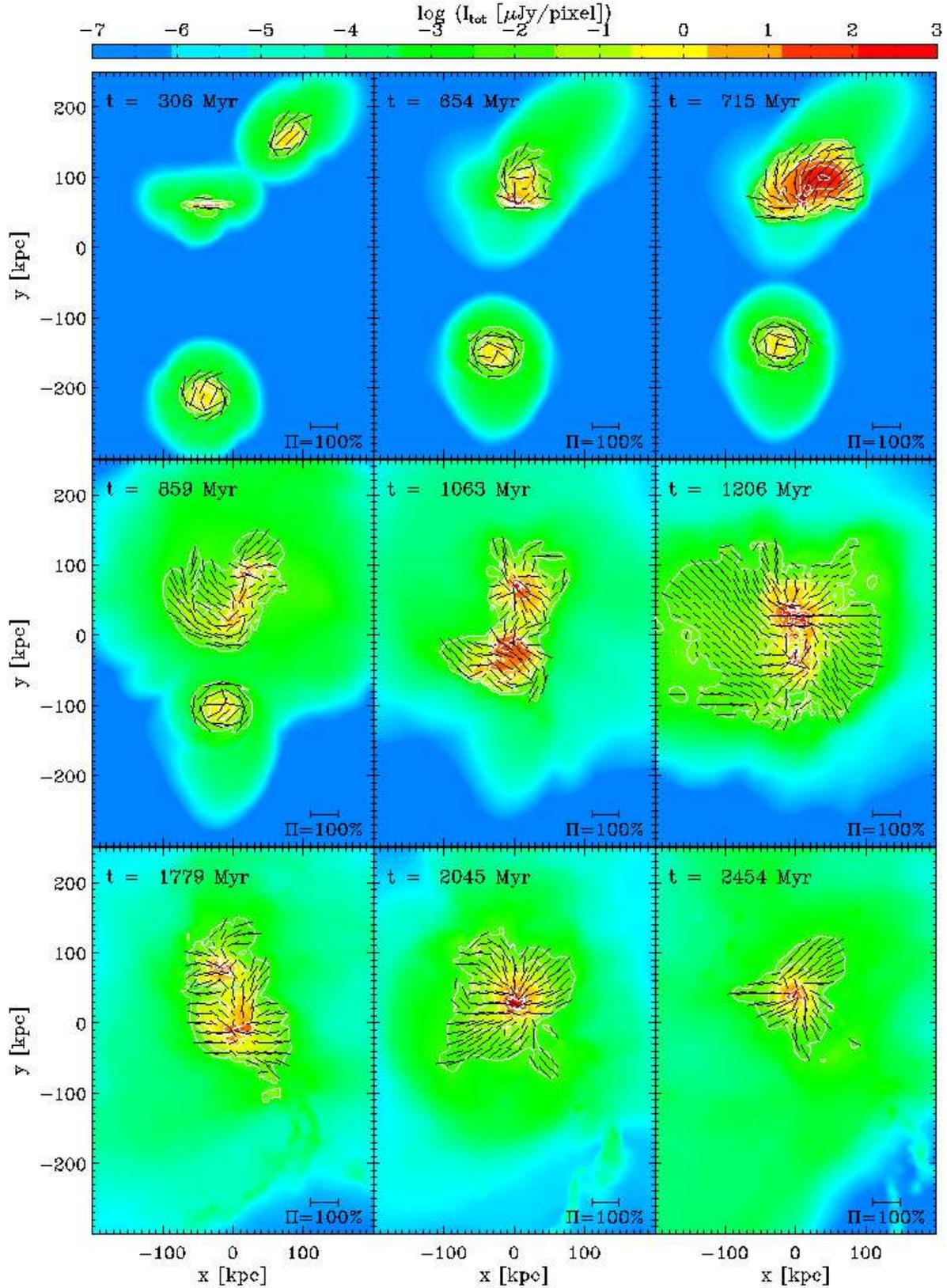, width=0.9\textwidth}
\end{center}
  \caption{Synthetic radio maps for the G6-IGM9 scenario at the same nine time steps as in Figs. \ref{evolution_bfld_G9-IGM12} - \ref{evolution_vrms_G9-IGM9}.  Colours visualize the total intensity (in $\mu$J/pixel). White contours show the polarized intensity, whereby the contour levels are 0.001, 0.01, 0.1, 1, 10 and 50 $\mu$J/pixel. Magnetic field lines derived from calculations of polarization are shown in black.
  \label{evolution_radio}}
\end{figure*}

Fig. \ref{evolution_radio} shows synthetic radio maps for the same nine time steps as in Figs. \ref{evolution_bfld_G9-IGM12} - \ref{evolution_vrms_G9-IGM9}. Colours visualize the total intensity (in $\mu$J/pixel). White contours show the polarized intensity, whereby the contour levels are 0.001, 0.01, 0.1, 1, 10 and 50 $\mu$J/pixel. The direction of the magnetic field is indicated by the black lines which are inclined by $\psi+\pi/2$, whereby the data of the 512$^2$ array has been rebinned to a 50$^2$ array, thus lowering the assumed angular resolution to $\approx13''$. The length of this lines is scaled according to the degree of polarization $\Pi_\mathrm{obs}$, the length-scale is given in the lower right corner of each plot. We show the magnetic field lines only where the polarized intensity is higher than 0.01 times the maximum polarized intensity in the beginning of the simulation, corresponding to a threshold of $\approx 0.001$ $\mu$Jy/pixel.

At the beginning of the simulation, the initially unidirectional magnetic field lines within the galaxies are wound up by the differential rotation of the disks. Hence, the magnetic field vectors derived from polarization are showing a nearly toroidal pattern at $t\approx300$ Myr (upper left panel). After the first collision (upper central panel), when shocks are driven into the IGM, the total and polarized intensities are considerably enhanced behind the shocks (upper right panel). Thereby, the magnetic field exhibits a regular pattern aligned with the direction of the outflow. In the subsequent evolution, prominent tidal arms are developing (middle left panel). These tidal arms are also visible in the total and polarized radio emission and are traced by the magnetic field lines. Presumably, the magnetic field lines have been stretched by the shear flows of the tidal structure. At time of and after the second collision (middle central and right panels), the polarized synchrotron emission is also visible outside the disks, showing that the interaction driven shocks and outflows have already magnetized parts of the IGM. During the subsequent evolution (lower left panel), the gas motions inside the galaxies become more random and thus the polarization of the radio emission generally decreases. At time of the final merger (lower central panel) shock driven gas flowing out of the merging system again gives rise to a high degree of polarization. By the end of the simulation, when most of the gas has been driven to the central core of the merged system (lower right panel), the total emission is concentrated around this core, and the polarization is tracing the weak outflows which are still driven into the IGM.

Fig. \ref{evolution_radio} does not reveal whether the polarized emission originates from an unidirectional or a reversing magnetic field configuration. Therefore, below we also present synthetic rotation measure ($RM$) maps of the simulated system. The value of $RM$ gives the strength by which the polarization vector of polarized radiation passing a magnetized plasma is rotated, whereby the rotation angle is $\phi=RM\cdot\lambda^2$. The $RM$ value for each simulated particle is calculated according to (cf. \citealp{Rybicki&Lightman1986})
\begin{equation}
  RM(\mathrm{particle})=\frac{e^3}{2\pi m_e^2 c^4}n_e B_\|,
\end{equation}
with $n_e$ the number density of thermal electrons (equal to the number density of thermal protons and thus proportional to the gas density) and $B_\|$ the magnetic field component along the line-of-sight. $RM$ is positive (negative) for a magnetic field directed toward (away from) the observer. The cumulative $RM$ is obtained by integrating along the line-of-sight, whereby individual $RM$ values may add or cancel. Thus, $RM$ distributions which show reversals on small scales indicate a reversing magnetic field. The cumulative $RM$ is binned on a grid as before in Fig. \ref{evolution_radio}.

\begin{figure*}
\begin{center}
  \epsfig{file=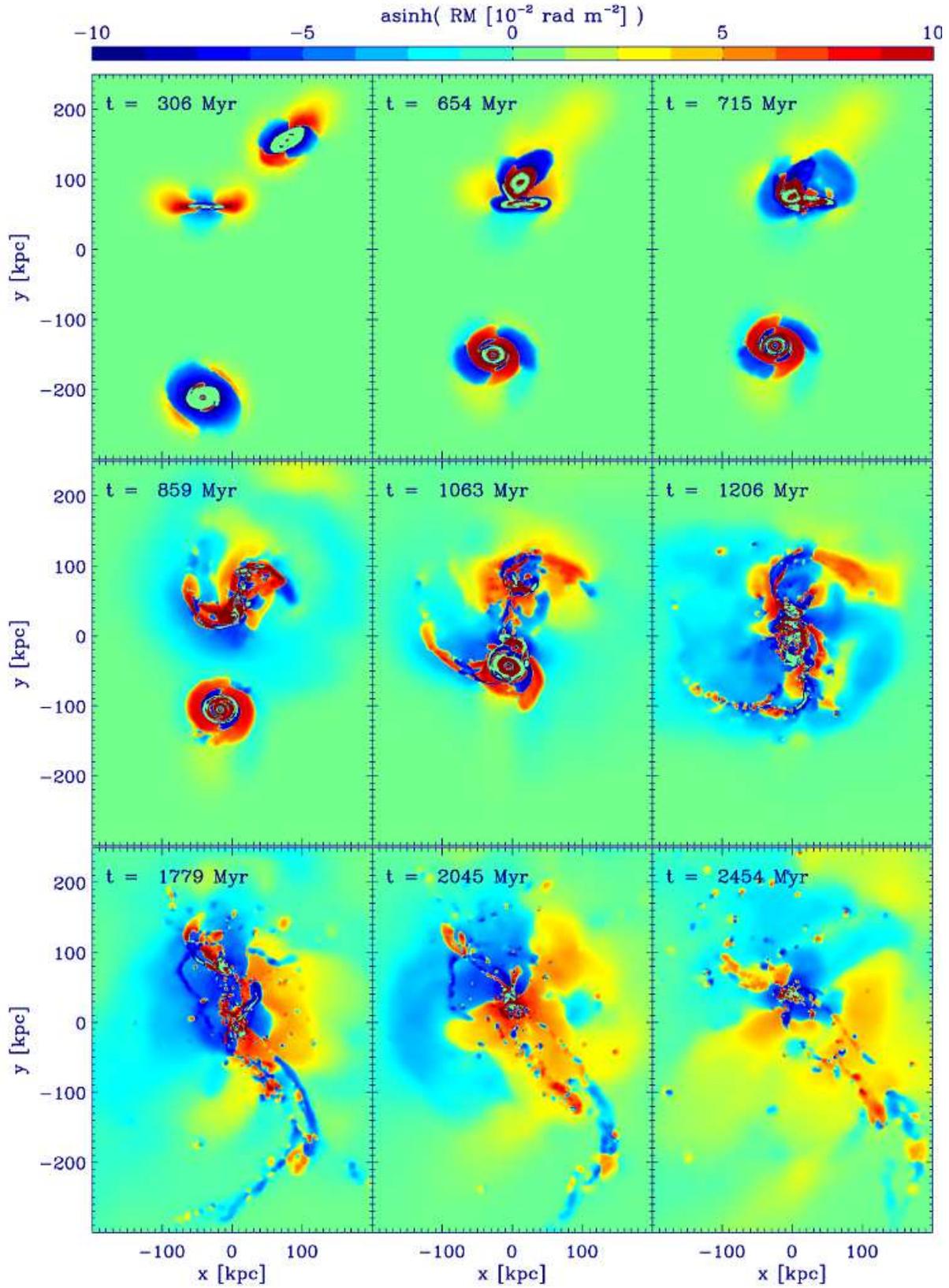, width=0.9\textwidth}
\end{center}
  \caption{RM maps for the G6-IGM9 scenario at the same nine time steps as in Fig. \ref{evolution_radio}. Colors visualize the RM value (in $10^{-2}$~rad~m$^{-2}$) on a asinh-scale for a better visibility of the small RM values in the IGM.
  \label{evolution_RM}}
\end{figure*}

Fig. \ref{evolution_RM} shows $RM$ maps for the G6-IGM9 scenario at the same nine time steps as in Fig. \ref{evolution_radio}. Colors visualize the $RM$ (in $10^{-2}$~rad~m$^{-2}$) on a asinh-scale for a better visibility of the small RM values in the IGM (caused by the low electron density in these regions). Note that on this scale values of 1, 5 and 10 correspond to $RM\approx0.01$, 0.74 and 110~rad~m$^{-2}$, respectively.

In the very beginning of the simulation ($t=0$~Myr, not shown), the lower galaxy (G3), which is seen exactly face-on, does not show any $RM$ because of the lack of a $B_\|$ component. The two other galaxies, which are inclined with respect to the line-of-sight, show a positive $RM$. The subsequent initial infall of IGM gas onto the galaxies and the winding of the initial magnetic field due to differential rotation lead to the development of the $RM$ patterns seen at $t=306$~Myr (upper left panel). E.g., the $RM$ is positive at both sides of the galaxy seen edge-on because the winding of the initially homogeneous magnetic field results in a field pattern with the field lines directed towards the observer at both sides of the galaxy. The lack of $RM$ in the central parts of the galaxies results from the cancelation of the (symmetric) contributions from both sides of the galaxies along the line-of-sight. Generally, the $RM$ does not show reversals on small scales, showing that the magnetic field is unidirectional on scales $\geq 10$~kpc. This regularity is a result of the homogeneity of the initial magnetic field. After the first collision ($t=654$~Myr), the $RM$ values within the regions of polarized emission (upper right panel, cf. Fig. \ref{evolution_radio}) are all negative, strongly indicating an unidirectional magnetic field configuration. This is also true for the interaction-driven outflows at later times (middle left to lower right panels), whereby the sign of the $RM$ may change depending on whether the outflow is driven predominantly towards or away from the observer. This large-scale homogeneity of the $RM$ distribution is most probably a relic of the homogeneous initial IGM magnetic field. Furthermore, the $RM$ tends to change sign on the edges of tidal arms (e.g. middle central panel, $t=1063$~Myr), showing that the magnetic field within the arms is predominantly of galactic origin in contrast to the IGM magnetic field. At later times, the $RM$ distribution within the galaxies and the tidal arms shows frequent reversals, originating in the reversing magnetic field tangled by interaction-driven turbulence.

In summary, when shocks and outflows are driven out of the galaxies or tidal structures are forming, a high amount of polarized emission can be expected due to the stretching of the magnetic field lines by the gas flows. Thereby, the magnetic field structure may be unidirectional or reversing depending on the precedent magnetic field. Generally, high synchrotron intensity corresponds to high density regions, except for periods of intensive shock ejection. During these periods, the magnetic field is enhanced by the shocks and transported from the galaxies into the IGM by interaction-driven outflows. \footnote{See http://www.usm.uni-muenchen.de/people/kotarba/public.html for a movie of the polarization and $RM$ maps of the G6-IGM9 scenario.}

Given the importance of shocks for the synthetic polarization in our simulations it is reasonable to ascribe observed prominent regular magnetic field structures like the magnetic field ridge observed in the Stephan's Quintet to shock activity. This assumption holds even better, as electrons are accelerated within shocks (which we do not model in our calculations), thus giving rise to an even higher synchrotron emissivity.

\section{Conclusions and Outlook}\label{CONCLUSION}

We have presented for the first time high resolution simulations of a merger of three disk galaxies within an ambient, magnetized IGM. We have studied three different models for the initial magnetic field strengths within the galaxies and the IGM, respectively, and compared these models to a simulation excluding magnetic fields. The initial magnetic field strength range form $10^{-12}$ G to $10^{-9}$ G within the IGM, and from $10^{-9}$ G to $10^{-6}$ G within the galaxies. We find that the magnetic field saturates at a value of several $\mu$G within the galaxies and at roughly $10^{-8}$ G within the IGM, independent of the initial magnetic field. This saturation levels correspond to equipartition between the magnetic and the turbulent pressure in the system. This result is in agreement with previous studies of \citet{meAnt2009}, who have presented simulations of the interaction of two disk galaxies, particularly the Antennae system. However, \citet{meAnt2009} did not include an ambient IGM, wherefore they could not study the behaviour of shock propagation within the IGM and its magnetization.
The simulations presented in this paper show that the shock propagation within the IGM is changed significantly depending on the initial magnetic field model. Thereby, the stronger the initial magnetic field, the stronger the shocks driven into the IGM. This result suggests that the shocks are supported by the magnetic pressure, resulting in higher Mach numbers in the presence of a strong magnetic field.

The main findings presented in this paper can be summarized as follows:

\begin{itemize}

\item The magnetic field within the galaxies grows to the equipartition value between the turbulent and magnetic energy density during the subsequent interactions between the merging galaxies. The IGM is magnetized by outflows and multiple shocks driven by the interactions up to nearly equipartition with the turbulent energy density within the IGM.

\item The final saturation value for the galaxies is of order of several $\mu$G, and the final IGM magnetic field strength has an average value of $10^{-8}$ G, independent of the applied initial magnetic field. As the setup of the system was chosen arbitrarily and the system was allowed to evolve freely in time, it is very interesting that the final magnetic field values are consistent with observations.

\item The growth of the IGM field is more efficient, the higher the galactic magnetic field, suggesting that magnetic energy is transported from the galaxies into the IGM.

\item The initial values of the magnetic field in the galaxies and in the IGM affect the propagation of shocks in the IGM: The stronger the initial field, the faster the shock propagation, suggesting that the shocks are gaining higher Mach numbers due to magnetic pressure support. This effect might be referred to as ``magnetic shock acceleration''.

\item The higher Mach numbers are also reflected in the higher temperatures of the shock heated IGM gas, and the shock-heated region is larger the stronger the shock.

\item Shocks play an important role for the polarized emission of an interacting system. Always when shocks are driven into the ambient IGM by an interaction, a high amount of polarized emission can be expected.

\end{itemize}

The presented simulations of a merger of three disk galaxies do not only provide insights in the evolution and significance of magnetic fields in a highly nonlinear environment, but also agree well with observations. The typical observed value of several $\mu$G in different types of local galaxies shows to be a typical value which arises naturally in merging systems. Also, the typical observational estimates for the upper limit of IGM magnetic fields of approximately $10^{-9}-10^{-8}$ G (\citealp{Kronberg2008} and references therein) are in agreement with the IGM magnetic field in our simulations.

Given the complexity of the simulated system, for a future project, it would be worthwhile to perform detailed shock simulations in a magnetized IGM setting. Parameter studies of the dependence of shock characteristics on the applied magnetic field strength and structure would lead to a deeper understanding of the evolution of magnetic fields during galactic interactions.

Finally, we emphasize that the efficient strengthening of magnetic fields during subsequent galaxy interactions up to levels consistent with observations is of particular interest within the framework of CDM clustering models. On the basis of the simulations presented in this paper and in \citet{meAnt2009}, we can assume that every galactic interaction contributes to the magnetization of the affected galaxies and the ambient IGM. As the phase of structure formation in the Universe is accompanied by frequent, subsequent interactions of galaxies and galactic subunits, the observed present-day magnetic fields might at least to some extent be the result of interaction-driven amplification processes in the early universe. Furthermore, the transport of magnetic energy from the galaxies into the IGM by interaction-driven outflows could explain the existence of IGM magnetic fields of order $10^{-9}$ - $10^{-8}$ G already at high redshifts. We note, however, that the simulations presented here picture the interaction of three fully evolved present-day galaxies, wherefore they can not be compared directly with interactions in the early universe. High redshift galaxies are known to be very different than present-day galaxies. However, subsequent galaxy interactions were more frequent in the early universe. Thus, it is reasonable to assume that the cumulative interaction-driven magnetic field amplification of even small initial seed fields might have magnetized young galaxies and their environment already at high redshifts. More studies of magnetic field evolution during the early phases of the Universe, preferably in a cosmological context, would thus be of particular interest for the whole astronomical community.

\section*{Acknowledgments}
We thank the referee Anvar Shukurov for his valuable comments and suggestions which helped us to improve this paper significantly.
This research was supported by the DFG Cluster of Excellence "Origin and Structure of the Universe" (www.universe-cluster.de). K.D. acknowledges the support by the DFG Priority Programme 1177.

\bibliographystyle{mn2e}

\appendix

\section{Appendix}\label{Appendix}

In Figs. \ref{evolution_bfld_G6-IGM9}, \ref{evolution_temp_G6-IGM9} and \ref{evolution_vrms_G6-IGM9} we showed the evolution of the magnetic field, the temperature and the rms velocity as a function of time for the G6-IGM9 scenario. Below, we show the corresponding figures for the G9-IGM9, G9-IGM12 and G0-IGM0 scenarios, respectively (see table \ref{tab0}).

\subsection{Magnetic fields}\label{Appendix_bfld}

\begin{figure*}
\begin{center}
  \epsfig{file=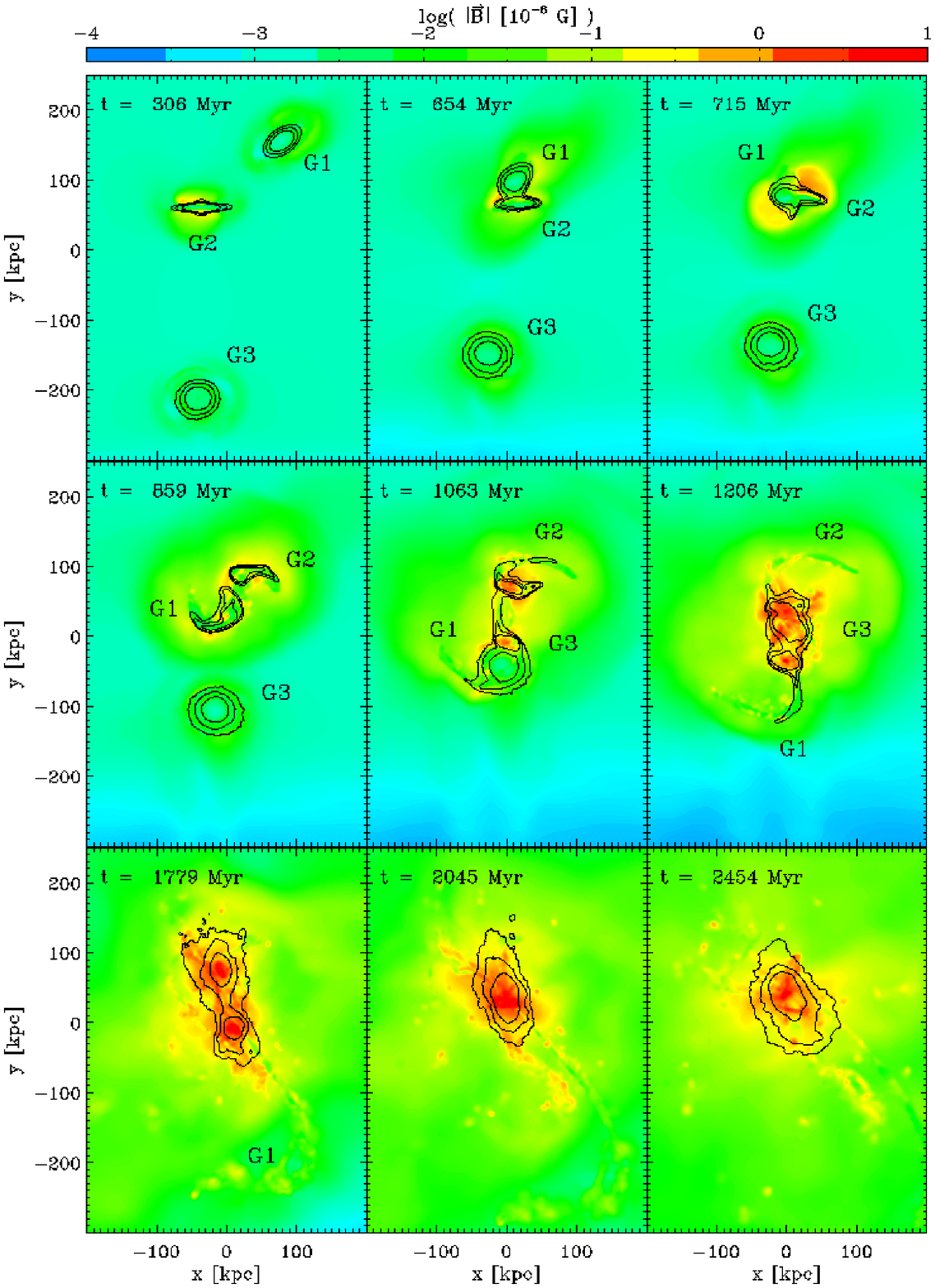, width=0.9\textwidth}
\end{center}
  \caption{Same as Fig. \ref{evolution_bfld_G6-IGM9}, but for the G9-IGM9 scenario.
  \label{evolution_bfld_G9-IGM9}}
\end{figure*}

\begin{figure*}
\begin{center}
  \epsfig{file=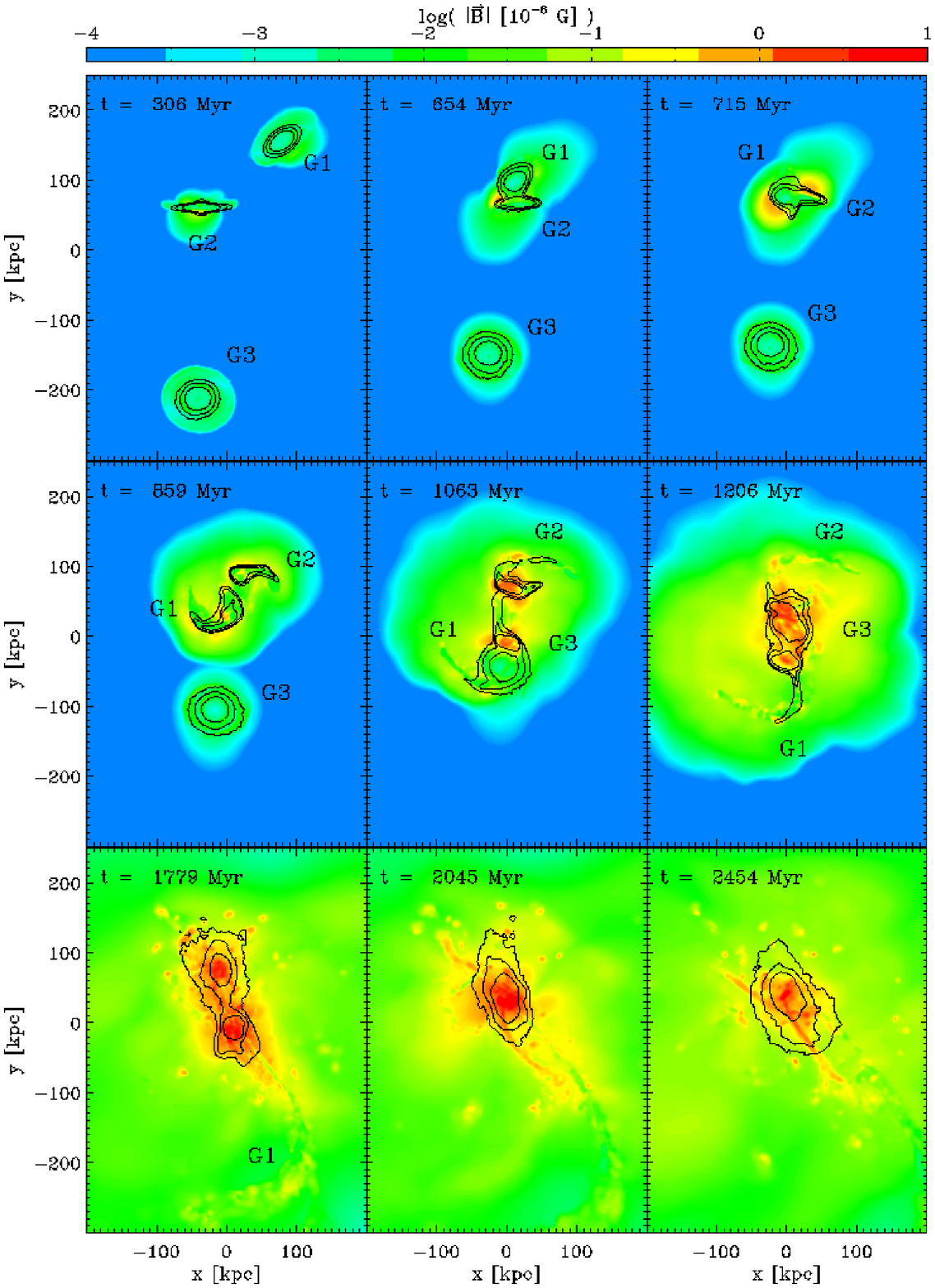, width=0.9\textwidth}
\end{center}
  \caption{Same as Fig. \ref{evolution_bfld_G6-IGM9}, but for the G9-IGM12 scenario.
  \label{evolution_bfld_G9-IGM12}}
\end{figure*}

Figs. \ref{evolution_bfld_G9-IGM9} and \ref{evolution_bfld_G9-IGM12} show the evolution of the mean line-of-sight magnetic field as a function of time similar to Fig. \ref{evolution_bfld_G6-IGM9}, but for the G9-IGM9 and G9-IGM12 scenarios, respectively. The final distributions and strengths of the magnetic fields are comparable within every scenario.

\subsection{Temperatures}\label{Appendix_temp}

\begin{figure*}
\begin{center}
  \epsfig{file=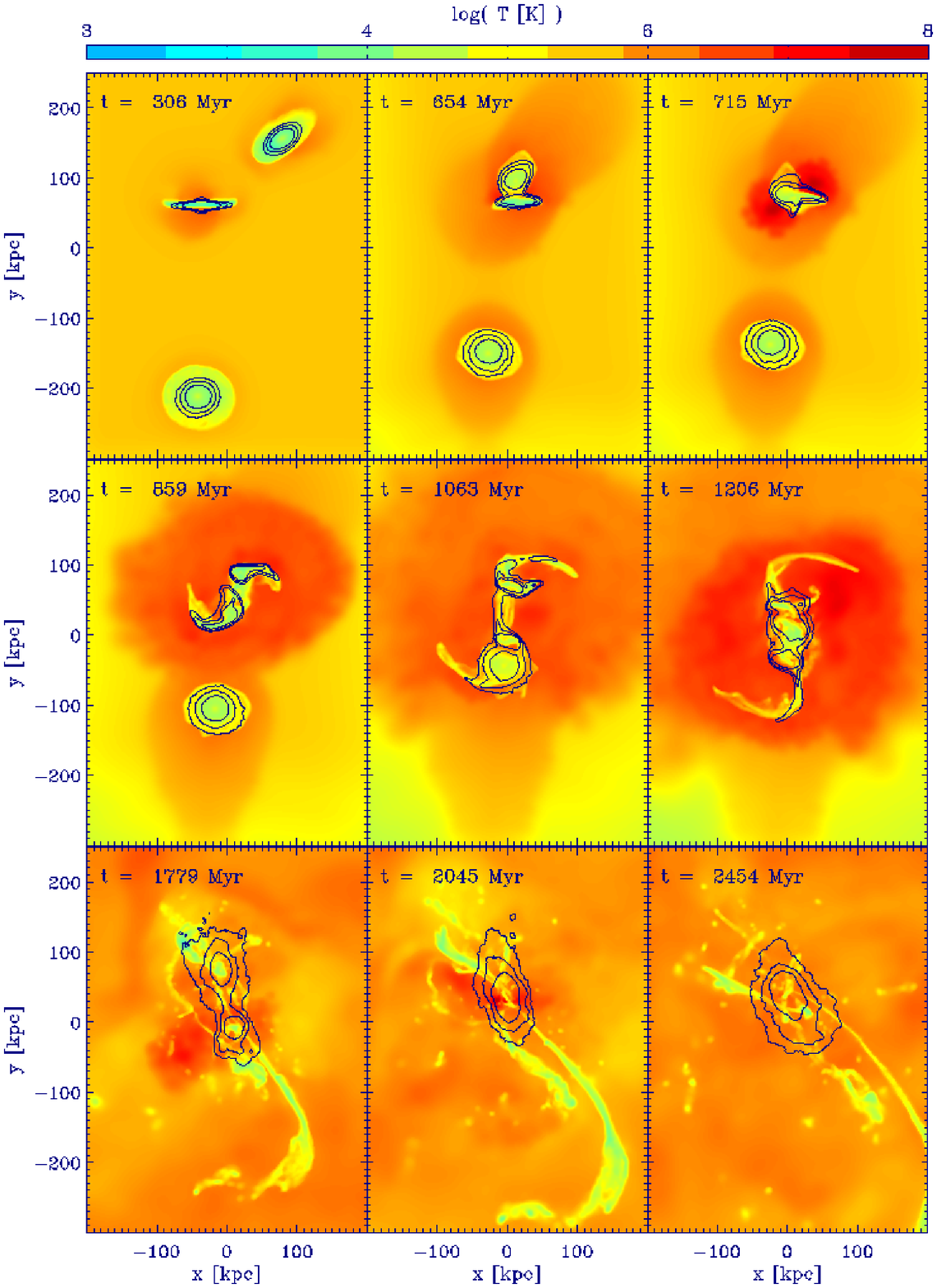, width=0.9\textwidth}
\end{center}
  \caption{Same as Fig. \ref{evolution_temp_G6-IGM9}, but for the G9-IGM9 scenario.
  \label{evolution_temp_G9-IGM9}}
\end{figure*}

\begin{figure*}
\begin{center}
  \epsfig{file=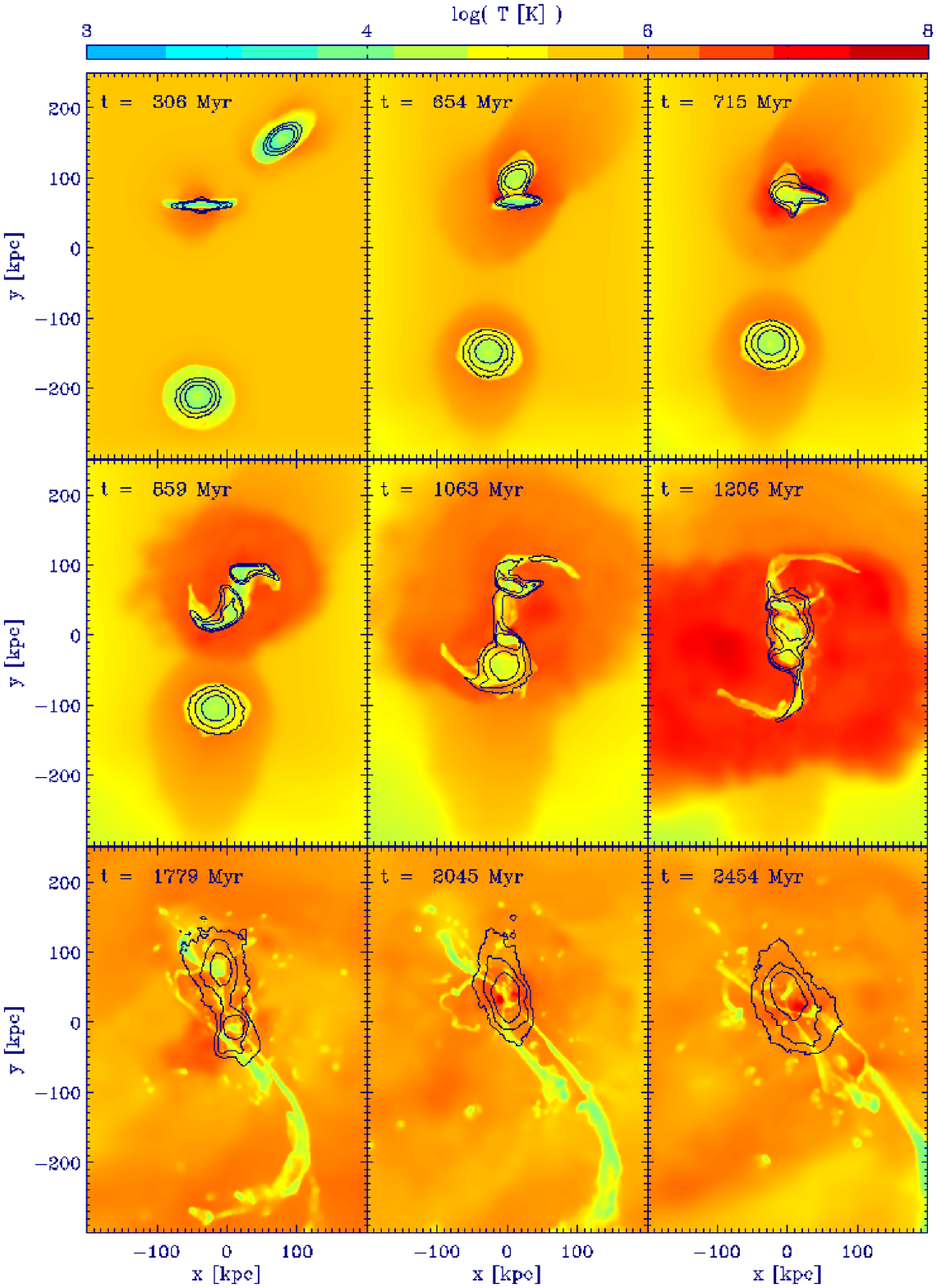, width=0.9\textwidth}
\end{center}
  \caption{Same as Fig. \ref{evolution_temp_G6-IGM9}, but for the G9-IGM12 scenario.
  \label{evolution_temp_G9-IGM12}}
\end{figure*}

\begin{figure*}
\begin{center}
  \epsfig{file=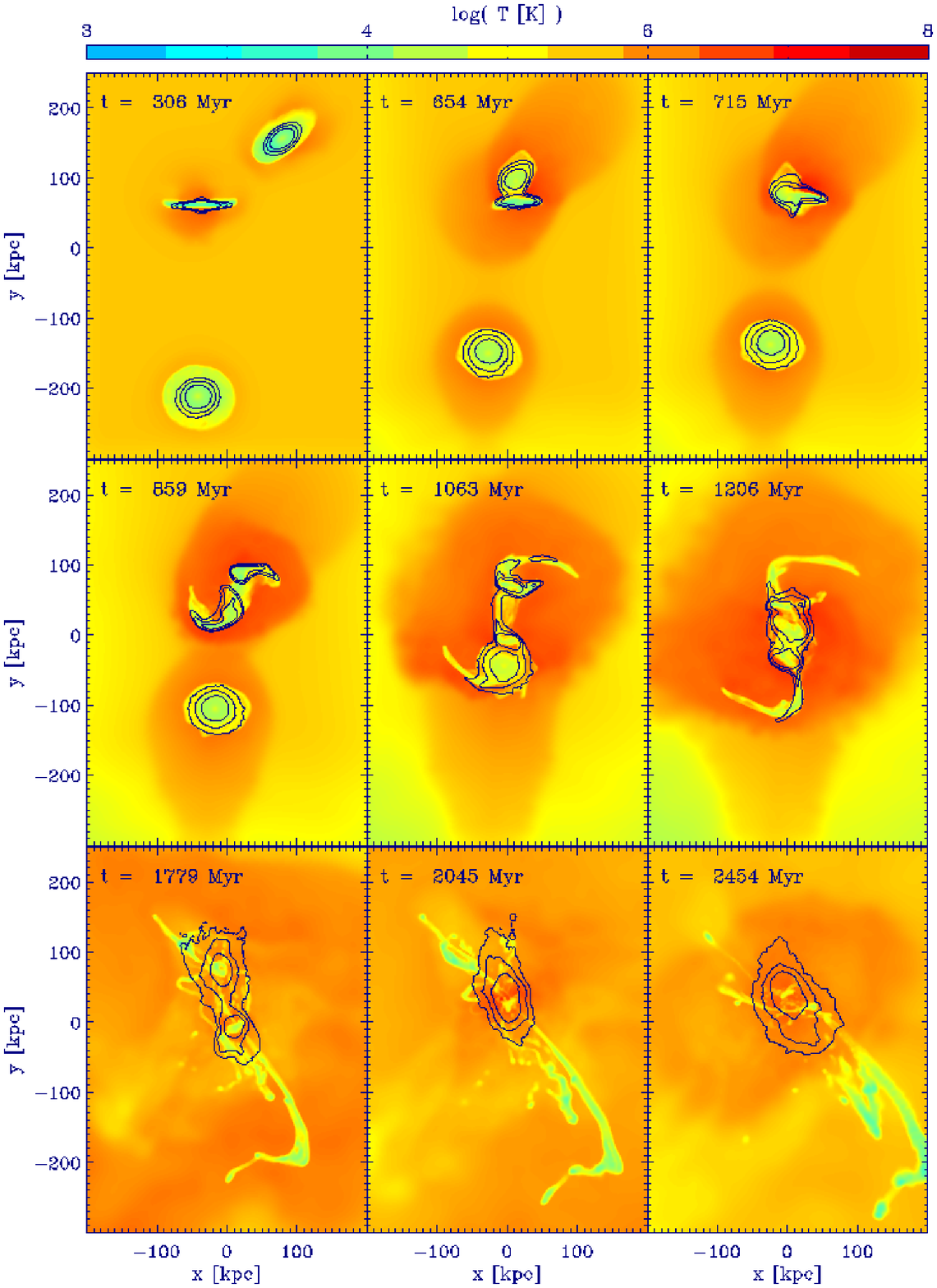, width=0.9\textwidth}
\end{center}
  \caption{Same as Fig. \ref{evolution_temp_G6-IGM9}, but for the G0-IGM0 scenario.
  \label{evolution_temp_G0-IGM0}}
\end{figure*}

Figs. \ref{evolution_temp_G9-IGM9}, \ref{evolution_temp_G9-IGM12} and \ref{evolution_temp_G0-IGM0} show the evolution of the mean line-of-sight temperature as a function of time similar to Fig. \ref{evolution_bfld_G6-IGM9}, but for the G9-IGM9, G9-IGM12 and G0-IGM0 scenarios, respectively. The stronger the shocks driven into the IGM, the higher the temperatures.

\subsection{RMS velocities}\label{Appendix_vrms}

\begin{figure*}
\begin{center}
  \epsfig{file=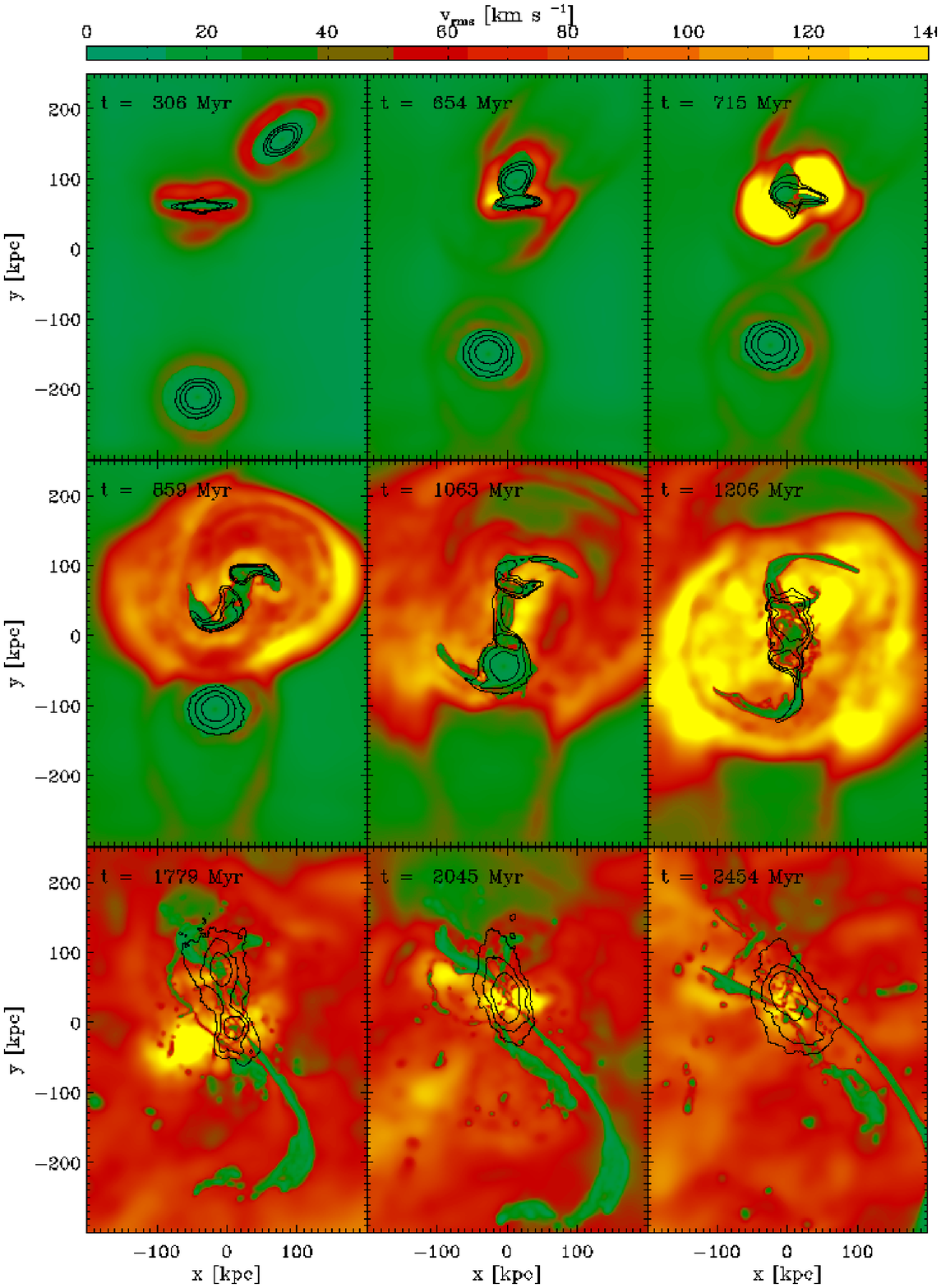, width=0.9\textwidth}
\end{center}
  \caption{Same as Fig. \ref{evolution_vrms_G6-IGM9}, but for the G9-IGM9 scenario.
  \label{evolution_vrms_G9-IGM9}}
\end{figure*}

\begin{figure*}
\begin{center}
  \epsfig{file=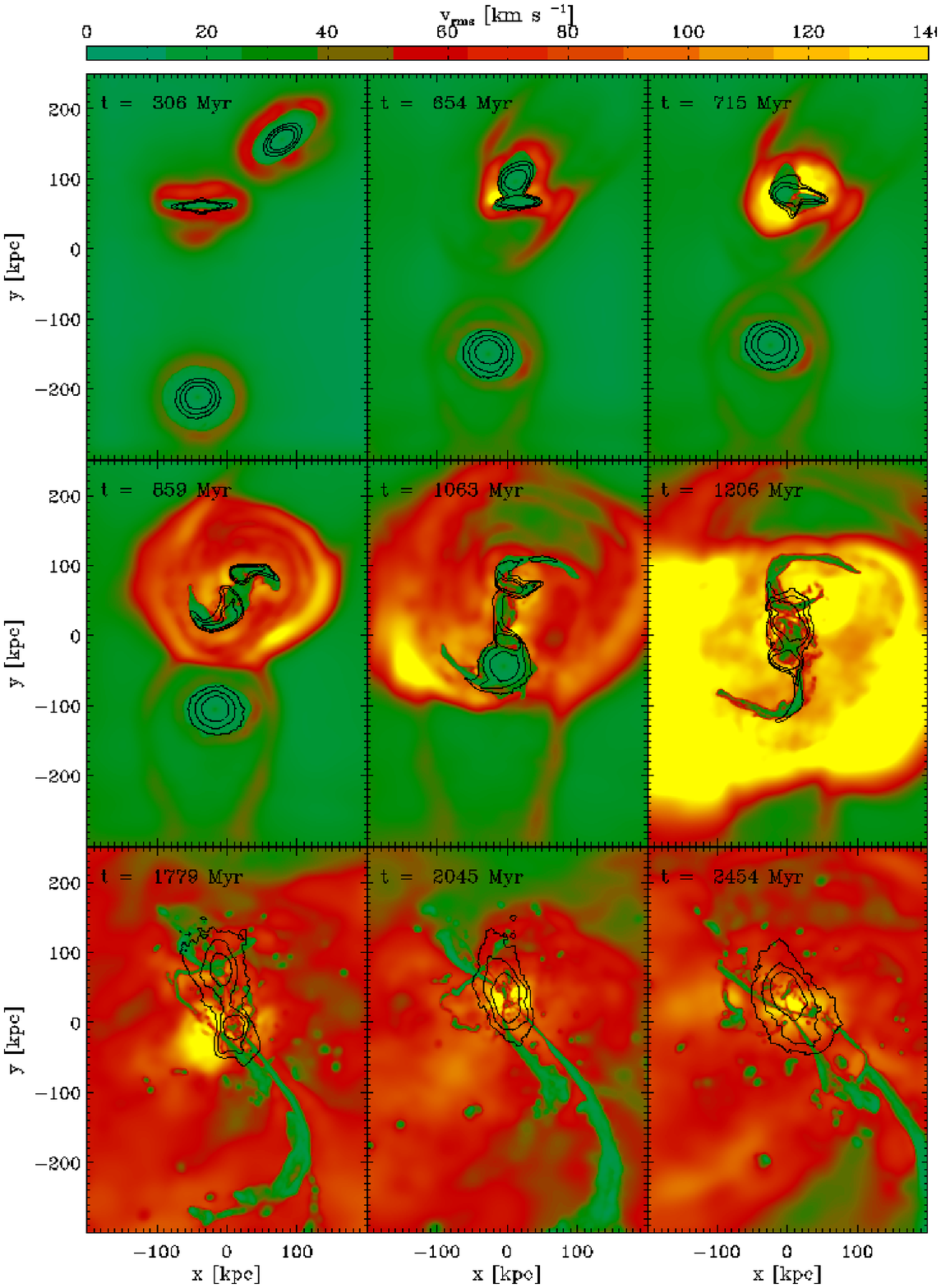, width=0.9\textwidth}
\end{center}
  \caption{Same as Fig. \ref{evolution_vrms_G6-IGM9}, but for the G9-IGM12 scenario.
  \label{evolution_vrms_G9-IGM12}}
\end{figure*}

Figs. \ref{evolution_vrms_G9-IGM9}, \ref{evolution_vrms_G9-IGM12} and \ref{evolution_temp_G0-IGM0} show the evolution of the mean line-of-sight rms velocity as a function of time similar to Fig. \ref{evolution_bfld_G6-IGM9}, but for the G9-IGM9, G9-IGM12 and G0-IGM0 scenarios, respectively. The higher the initial magnetic field, the faster the shock propagation within the IGM.

\begin{figure*}
\begin{center}
  \epsfig{file=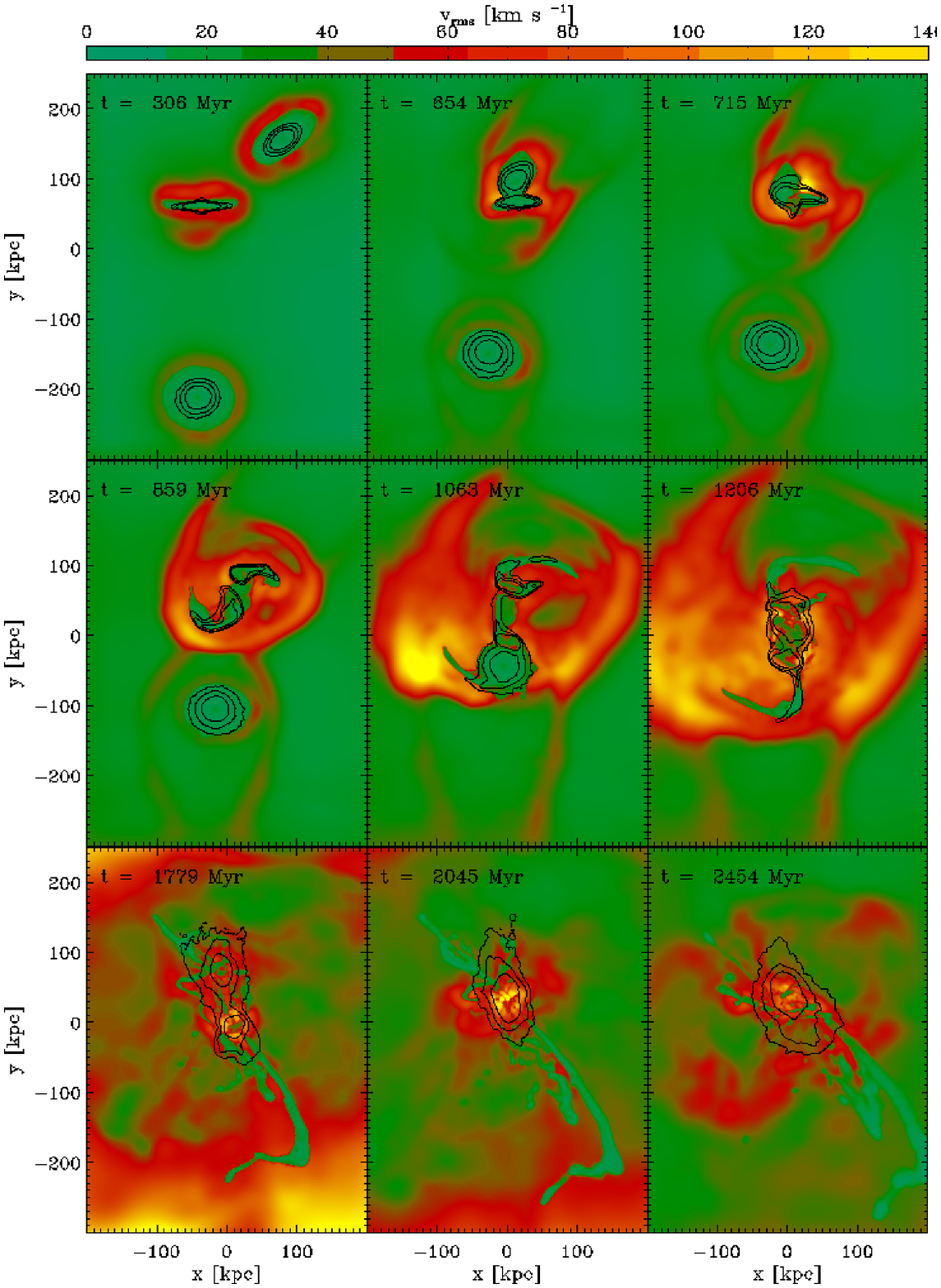, width=0.9\textwidth}
\end{center}
  \caption{Same as Fig. \ref{evolution_vrms_G6-IGM9}, but for the G0-IGM0 scenario.
  \label{evolution_vrms_G0-IGM0}}
\end{figure*}

\label{lastpage}
\end{document}